\documentclass[fleqn,usenatbib]{mnras}
\usepackage{newtxtext,newtxmath}

\usepackage[T1]{fontenc}

\DeclareRobustCommand{\VAN}[3]{#2}
\let\VANthebibliography\thebibliography
\def\thebibliography{\DeclareRobustCommand{\VAN}[3]{##3}\VANthebibliography}

\usepackage{graphicx}	
\usepackage{amsmath}	
\usepackage{float}
\usepackage[normalem]{ulem}
\usepackage{color}
\usepackage{xcolor}

\def\swift{\emph{Swift}}

\def\itpeak{\ensuremath{I_{T,\mathrm{peak}}}}
\def\impeak{\ensuremath{I_{M,\mathrm{peak}}}}

\def\ctfull{\ensuremath{C_{T,\mathrm{full}}}}
\def\ctpeak{\ensuremath{C_{T,\mathrm{peak}}}}
\def\cmpeak{\ensuremath{C_{M,\mathrm{peak}}}}
\def\cmfull{\ensuremath{C_{M,\mathrm{full}}}}
\def\efull{\ensuremath{E_{\mathrm{full}}}}
\def\epeak{\ensuremath{E_{\mathrm{peak}}}}

\def\pnew{\ensuremath{P_\mathrm{new}}}

\newcommand{\tim}[1]{\ensuremath{\times 10^{#1}}}

\title[Optimising transient discovery with \swift-XRT]{Optimising transient discovery with \swift-XRT }

\author[S. Srivastava et al.]{
S. Srivastava,$^{1}$\thanks{E-mail: ss1606@leicester.ac.uk}
P. A. Evans,$^{1}$
M. R. Goad$^{1}$
and R. A. J. Eyles-Ferris$^{1}$
\\
$^{1}$School of Physics \&\ Astronomy, University of Leicester, University Road, Leicester, LE1 7HR, UK\\
}

\date{Accepted XXX. Received YYY; in original form ZZZ}

\pubyear{\the\year{}}

\begin{document}
\label{firstpage}
\pagerange{\pageref{firstpage}--\pageref{lastpage}}
\maketitle

\begin{abstract}
The Living \swift-XRT Point Source Catalogue (LSXPS) enables near real-time searches for X-ray transients. Many detected candidates are faint, often near the XRT detection limit, and are classed as ``low significance,'' as it is often unclear whether their apparent brightening reflects a genuine transient or a statistical fluctuation. Some of these sources are affected by Eddington bias, a statistical effect that inflates measured fluxes near the detection threshold. We present a simulation-based Bayesian framework that corrects for this bias and provides more accurate probabilities for each source being truly transient, i.e. that its true intensity exceeds the historical 3$\sigma$ upper limit. Applied to LSXPS data, this method yields more reliable classifications, recovering over 500 transients above this threshold---more than an eight-fold increase over the original confirmed sample. Using extensive simulations based on real \textit{Swift}-XRT images, we validate the robustness of this approach, showing that it remains stable across varying exposure times and background conditions. These results demonstrate that the LSXPS transient probabilities, corrected for Eddington bias, provide a reliable and internally consistent framework for real-time X-ray transient identification.
\end{abstract}

\begin{keywords}
Catalogues -- Astronomical data bases -- Methods: statistical -- Transients -- X-rays: general -- Astronomical instrumentation, methods, and techniques
\end{keywords}



\section{Introduction}
\label{sec:intro}

X-ray astronomy allows us to probe some of the universe's most energetic and dynamic processes, including accretion onto compact objects, relativistic jets, and explosive transients. \textit{The Neil Gehrels Swift Observatory} (\swift), launched in 2004 to study gamma-ray bursts (GRBs), was equipped with rapid slewing, flexible scheduling, and \swift \ Target of Opportunity (ToO) response capabilities. While these features were tailored for GRB follow-up, they have made \swift\ an ideal platform for broader Time-Domain and Multi-Messenger (TDAMM) astrophysics \citep{Gehrels2004, Page2020,keivani2021swift}.

Beyond GRBs, \swift \ has uncovered a diverse range of high-energy transients. The Burst Alert Telescope (BAT; \citealt{BarthelmyBAT}) has revealed extreme phenomena such as ultra-luminous X-ray sources (ULXs) and jetted tidal disruption events (TDEs; \citealt{Burrows2011, Cenko2012, Bloom2011Sci, Brown2015MNRAS, Pasham2015ApJ,2011Sci...333..199L}). Transients can also be found among the objects serendipitously observed by the X-ray Telescope (XRT; \citealt{Burrows2005}). A notable example is SN2008D, a core-collapse supernova discovered during \swift-XRT observations of NGC 2770. These observations were part of a ToO programme, during which the new supernova and its accompanying X-ray transient were captured serendipitously \citep{Soderberg2008}.
X-ray transients span a wide variety of phenomena, including fast X-ray transients (FXTs) detected by missions like \emph{BeppoSAX} and \emph{ROSAT} \citep{Heise2001}, luminous flares from previously undetected sources in \emph{Chandra} archival data \citep{Jonker2013, Bauer2017, Quirola2022, quirola2023extragalactic,quirola2024probing}, and transients first detected and later identified in \emph{XMM-Newton} observations, including supernova shock breakouts and other previously unclassified X-ray transients \citep{AlpLarsson2020ApJ, DeLuca2021AA}; more recently, \emph{Einstein Probe} has revealed soft X-ray flares, FXTs and candidate TDEs \citep{Yuan2022,eyles2025kangaroo, 2025arXiv250925877L, 2025ApJ...990L..29S, 2025ApJ...994L..17O, 2025NatAs...9.1375L, 2026MNRAS.545f2064Q,2025ApJ...988L..13R,2026MNRAS.545f2021J,2025arXiv250417034L,2025SCPMA..6819511Z}. Many events found in archival \emph{Chandra} and \emph{XMM-Newton} data were discovered serendipitously during archival investigations of pointed X-ray observations. Large catalogues such as the \emph{Chandra} Source Catalog (CSC; \citealt{Evans2010}), \emph{XMM-Newton}'s 4XMM catalog \citep{Webb2020}, \emph{eRASS}-DE DR1\citep{Merloni2024eRASS1DR1}, and the \swift\ 2SXPS catalogue \citep{Evans2020} have enabled systematic searches for transient behaviour among detected sources. These have revealed previously unclassified or rare events, including TDEs, extreme flaring from active galactic nuclei (AGNs), and Galactic X-ray binaries, many of which had been missed during initial observations, \citep[e.g.][]{Jonker2013, eyles2025nine, 2021MNRAS.508.3820S}. However, these catalogues are inherently static, and take considerable time and effort to produce, meaning that the data they contain are often months to years old. As a result, many transients hidden in these catalogues are identified only after their occurrence, often limiting the possibility of rapid follow-up (e.g. \citealt{Miniutti2019}; \citealt{Giustini2020}; \citealt{Starling2011}; \citealt{Strotjohann2016}), although some sources have been recognised quickly enough to enable prompt observations, as in the case of the \emph{Chandra} transient CDF-S XT1 \citep{Bauer2017CDF-SXT1}. To overcome these limitations, the Living \swift-XRT Point Source Catalogue (LSXPS) was developed \citep{Evans2022}.

LSXPS represents a paradigm shift in X-ray source cataloging by introducing a continuously updated, real-time system for transient detection. Unlike its predecessors, LSXPS automatically incorporates new \swift-XRT data as it becomes available, allowing for the rapid identification of previously unknown X-ray transients. This dynamic approach not only enhances the efficiency of transient follow-up but also enables prompt multi-wavelength observations, providing valuable insights into the variability and evolution of known sources. A key demonstration of LSXPS’s capabilities came with the discovery of \swift~J023017.0+283603 (hereafter \swift~J0230), a unique transient identified on 2022 June 23, during a routine \swift-XRT observation of SN2021afkk \citep{Evans2023,Guolo2024}. Within three minutes of the data becoming available (approximately five hours after the observation), LSXPS flagged the source as a transient. A ToO observation was scheduled within an hour. Follow-up studies revealed \swift~J0230 to be an exceptionally soft X-ray source, exhibiting a blackbody-like thermal spectrum and quasi-periodic eruptions (QPEs) with a period of $\sim$ 25 days — significantly longer than most known QPEs. While the event was initially interpreted as a repeating partial TDE involving an intermediate-mass black hole (IMBH), alternative interpretations have since emerged. A recent analysis by \citet{Pasham2024} suggests that the eruptions may result from a planet being periodically stripped rather than a main-sequence star. This is consistent with the proposed model by \citet{Guolo2024}, who also advocate a planetary origin based on energetics and recurrence time. The discovery of \swift~J0230 highlights the scientific value of real-time, sensitive X-ray detection systems, and demonstrates LSXPS’s potential to uncover new classes of transient astrophysical phenomena.

To identify new transients, the LSXPS pipeline follows a structured process. First, all X-ray sources in newly-downlinked \swift-XRT data are detected and positionally cross-matched against existing catalogues, including LSXPS itself, to identify uncatalogued sources. For each of these, the pipeline searches archival data from \swift, \emph{XMM-Newton}, and \emph{ROSAT} \footnote{eRASS1 will be added soon.} to determine the deepest available $3\sigma$ upper limit on the historical flux. This limit
is then compared to the peak flux measured in the new detection. If the $1\sigma$ lower bound of this peak flux exceeds the historical upper limit, the source is flagged as a candidate transient. These candidates
are then reviewed manually by the LSXPS team and classified using a set of empirical criteria described in \cite{Evans2022}\footnote{The classification summaries are also available at \url{https://www.swift.ac.uk/LSXPS/docs.php\#transient}.}.

It is important to note that source detection always uses the full dataset available at analysis time\footnote{\swift\ data are released immediately, without waiting for observations to be complete, meaning that a given observation will often be analysed multiple times, with increasing exposure.}, whereas the peak intensity is determined separately, by measuring the source intensity both over the full dataset and in each individual `snapshot'\footnote{Roughly defined as a period of contiguous exposure; for \swift's low-Earth orbit a single snapshot is at most $\sim$2.7~ks.}, and taking whichever value has the highest $1\sigma$ lower limit as the peak. In practice the peak is therefore frequently set by a single snapshot rather than the full observation. As a consequence, the peak intensity of a transient candidate \emph{and the exposure time over which it is measured} may both differ from, and be considerably shorter than, the average intensity and total exposure over which the source was detected. For full details of the LSXPS transient detection pipeline, see \cite{Evans2022}, especially their Fig.~1 and Sections~2 and~4.

A key challenge in transient detection lies in accurately distinguishing genuine astrophysical events from statistical fluctuations — particularly for low-significance detections near the detector's sensitivity limit. The distinction between a `transient' and an `outburst' is itself inherently subjective, as a transient may simply be an outburst from a source whose quiescent flux lies below historical detection thresholds: how far below is often unknowable. Furthermore, low-count detections ($<\sim30$ counts) are particularly vulnerable to Eddington bias in which, due to the asymmetry of the Poisson distribution and the fact that fainter sources are more populous, the intensity of sources near the detection threshold is systematically overestimated \citep{Eddington1940}.

For \swift-XRT, this effect can be seen for sources from which fewer than 30 counts were detected, and can lead to a factor of $\sim$five overestimate in count-rate; see \citet{Evans2014}, their fig.~10. As a result, some transient candidates identified by LSXPS will in reality have fluxes below the historical upper limit and thus show no real evidence of transient nature, creating uncertainty in population studies and hindering the robust identification of genuinely new transients. To mitigate this, the LSXPS pipeline includes a `low-significance' category for sources with fewer than 30 counts and whose peak count rate is less than 3-$\sigma$ above the historical 3-$\sigma$ upper limit. This conservative classification accounts for approximately 54\% of the LSXPS transient candidates in our sample, and reflects the challenge of confidently validating low-count transients. The framework developed in this work directly addresses this ambiguity by providing a statistically rigorous method to estimate the probability that a source’s true flux exceeds the historical upper limit, fully incorporating detector effects and statistical biases. We present a simulation-based Bayesian approach that accounts for the detection capabilities of \swift-XRT, distribution of background source fluxes, and Poisson noise. This provides a robust probabilistic classifier for transient candidates and enables improved separation between transient events and those arising from Poisson noise / Eddington bias.

\section{Methods}

To establish whether a newly discovered X-ray source in LSXPS is a genuine transient, we must ask:  
\emph{given the LSXPS-detected source has a measured peak intensity \impeak, what is the probability that it is in reality brighter than the historical 3-$\sigma$ upper limit?} i.e.\  we must calculate $P(\itpeak>L | \impeak)$ where $\itpeak$ is the true peak intensity and $L$ is the historical upper limit.

Our method can be summarised as follows:

\begin{enumerate}
    \item Generate datasets containing simulated sources of known intensity \( I_T \).
    \item Process these datasets through the LSXPS pipeline, determining for each whether the source was detected, and if so, at what measured intensity \( I_M \).
    \item Apply Bayes' theorem, accounting for the intrinsic distribution of source intensities and the XRT detection probability, to derive the posterior probability distribution of the true intensity $P(\itpeak | \impeak)$ and hence determine $P(\itpeak>L | \impeak)$.
\end{enumerate}

While source intensity, $I$ (count rate) is the relevant property for determining whether or not a source is a transient, in our simulations it is often necessary to think in terms of the integrated counts, $C= I E$, where $E$ is the exposure time over which $I$ was measured.

\subsection{Generating simulated images} 

The first step of our analysis involved generating synthetic \swift-XRT images containing simulated point sources. These simulations were designed to span the full range of true source intensities, \emph{$I_T$}, that could potentially produce a `low-significance' transient candidate. For the lower bound, we identified the minimum intensity, $I_{T,\mathrm{min}}$, such that
recording 2 counts is a ${\sim}4\sigma$-rare event, i.e.\
$P(C = 2 \mid I_T) \approx 5\times10^{-5}$ in the simulated exposure.
For a Poisson mean $\mu = I_T E$ this corresponds to $\mu = 0.01$,
giving $I_{T,\mathrm{min}} = 0.01/E$~cts~s$^{-1}$. The upper bound, $I_{T,\max}$, was set to $0.6$\,cts\,s$^{-1}$ since above this level pile-up becomes significant for XRT, making our simulations more complex; however, even in the shortest XRT exposures, such sources would yield $\gg 30$ counts and so would not be classified as `low-significance'. This intensity is also more than two orders of magnitude above the sensitivity threshold (and hence upper limits) of the \emph{ROSAT} All-Sky Survey \citep{Voges1999,Boller2016} and the \emph{XMM-Newton} Slew Survey \citep{Saxton2008,Saxton2011}, using a count-rate-to-flux conversion based on a typical AGN spectrum with photon index $\Gamma = 1.7$ and hydrogen column density $N_{\rm H} = 3 \times 10^{20}\,\mathrm{cm}^{-2}$ \citep{Evans2014}.

\begin{figure}
    \centering
    \includegraphics[width=\linewidth]{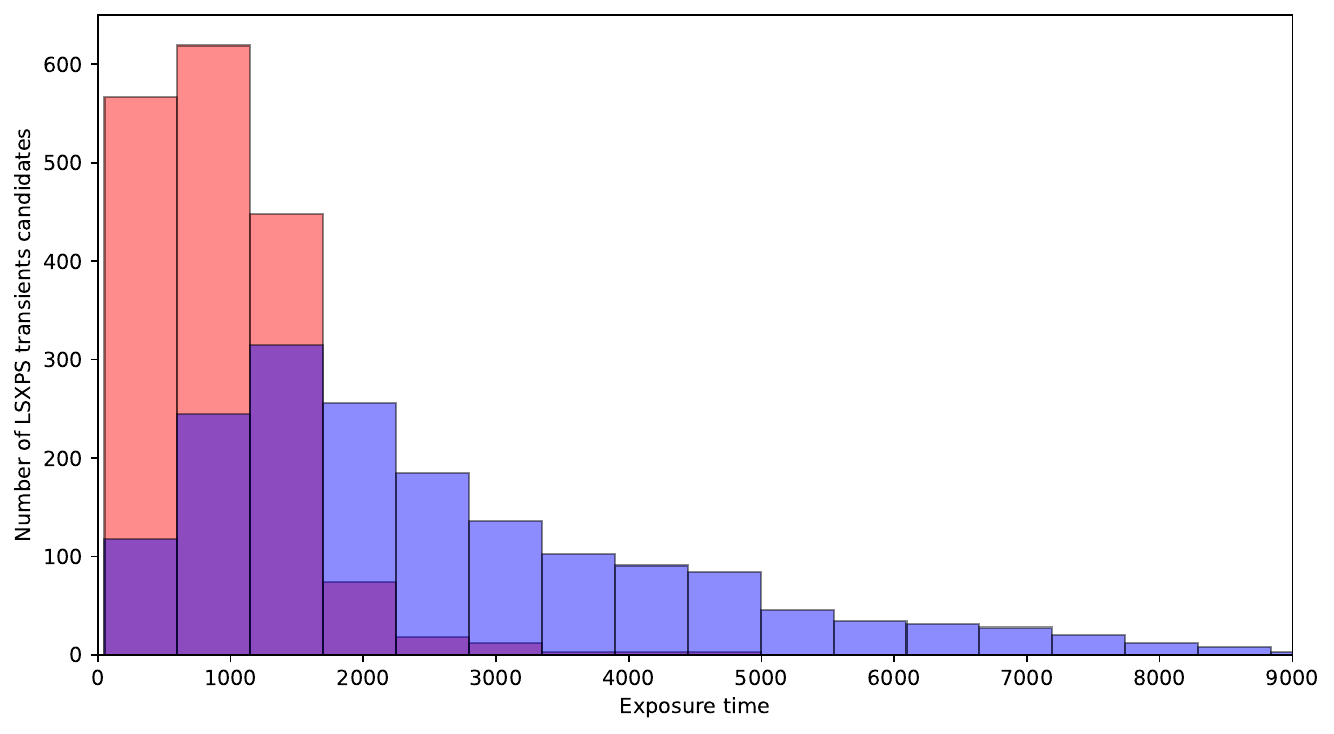}
    \caption{ The distribution of exposure times over which \impeak\ was measured for the LSXPS transient candidates
    in our sample (red) and in which the source was first detected (blue).}
    \label{fig:exposure}
\end{figure}

To determine an appropriate exposure time for the simulations, we examined the distribution of exposure peak values for fields containing transients in LSXPS (Figure. \ref {fig:exposure}). Based on the distribution, we selected  2.0 ks for our initial experiments.

We simulated 85 distinct source intensities, spaced uniformly in logarithmic space between the limits identified above. This sampling is dense enough to ensure a smooth posterior $P(T \mid M)$, the
probability of the true brightness $T$ given the measured brightness $M$ (defined fully in Section~\ref{sec:results}), while still allowing a rapid, real-time application of our results to new transients.

 For each intensity level, 30,000 independent simulations were performed sufficient to yield an average of approximately 2 sources at the $4\sigma$ extreme of the Poisson distribution i.e.\ to ensure that we sample probability space sufficiently.

Each simulated image was constructed by placing a point source at a random position within a real \swift-XRT sky field . The number of photons for each source was drawn from a Poisson distribution with a mean equal to the product $C_T = I_T E$ -- i.e.\ we treat the sources as being of constant flux during the observation, rather than deploying some arbitrary variability model. We return to this point in Section \ref{sec:Det}.

\begin{figure}
    \centering
    \includegraphics[width=\linewidth]{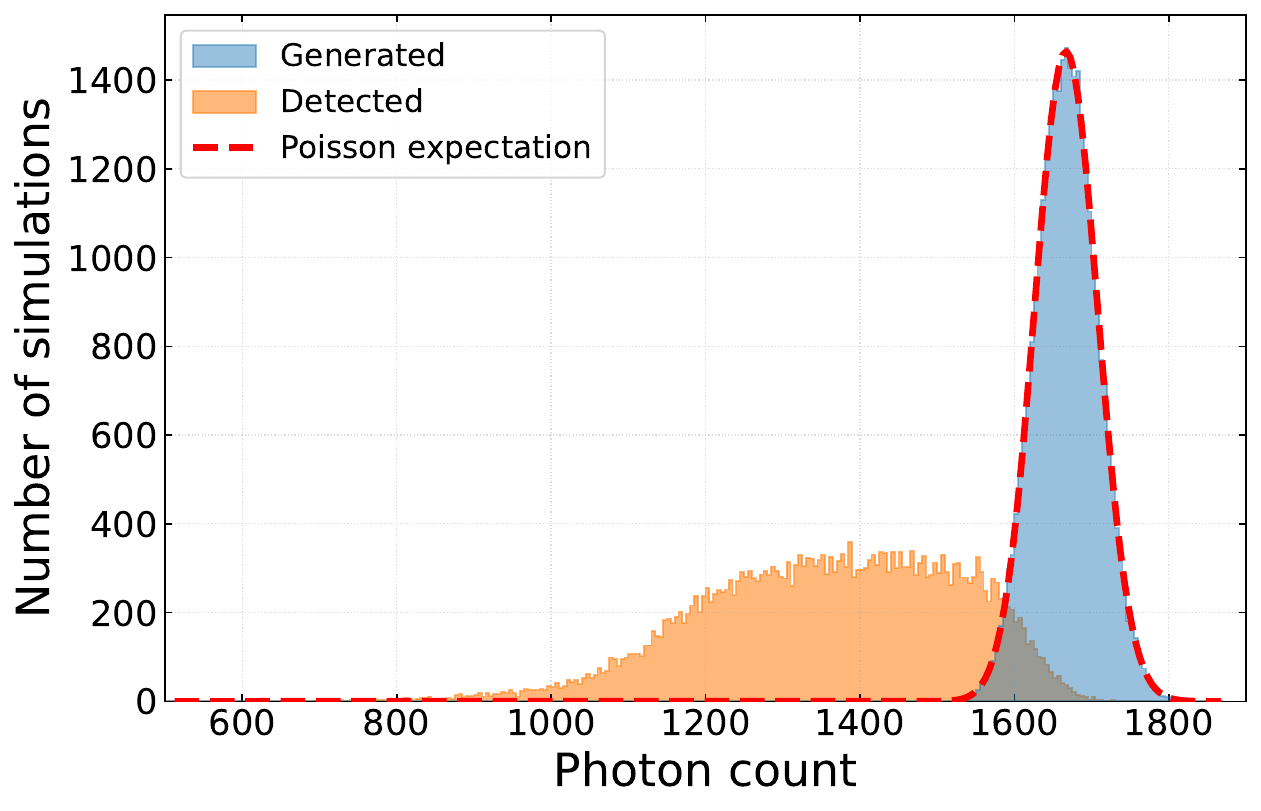}
    \caption{Distribution of the number of photons generated for a source of intensity $I_T = 0.6$~ct~s$^{-1}$ (with the expected Poisson distribution overplotted in red, demonstrating consistency with the theoretical expectation), shown together with the number of photons actually added to the image after applying the \swift-XRT exposure map. The difference between the two reflects the impact of vignetting and
dead pixels.}
    \label{fig:gp}
\end{figure}

The spatial distribution of photons was drawn from the instrument’s point spread function (PSF).  To account for vignetting and bad pixels, we then folded the photons through the exposure map: for a given detector pixel we define $t_{\text{pixel}}$ as the effective exposure time from the exposure map and $t_{\text{nominal}}$ as the nominal on-axis exposure time of the observation. Each photon landing in that pixel was retained with probability $t_{\text{pixel}}/t_{\text{nominal}}$.
\citep{Moretti2005, SwiftXRTCALDB2018}. Fig. \ref{fig:gp} illustrates the effect of applying the exposure map by comparing the number of photons initially generated with the number retained after vignetting and bad-pixel corrections.

\subsection{Processing the simulated images}

\begin{figure}
    \centering
    \includegraphics[width=\linewidth]{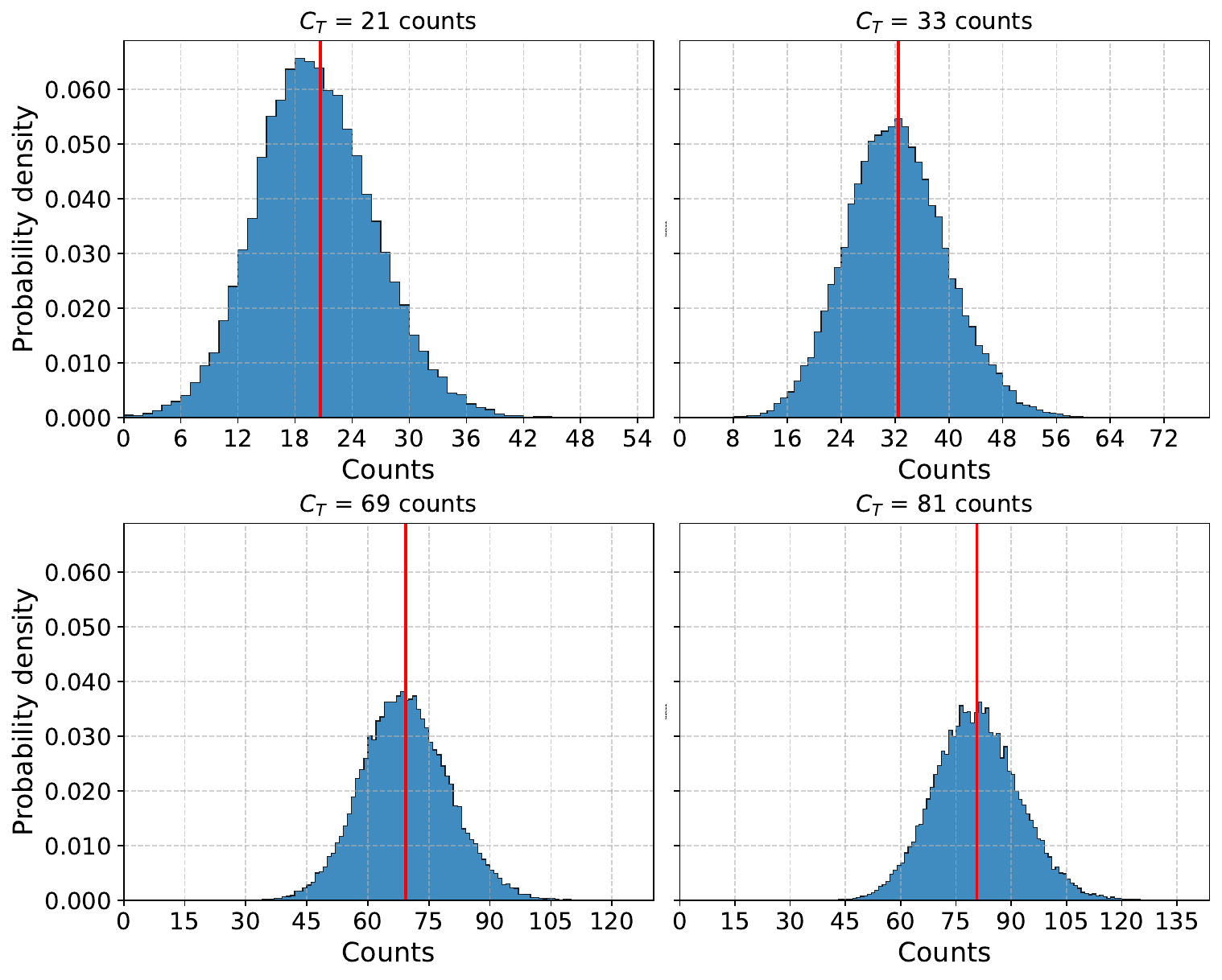}
     \caption{Examples of the probability distributions $P(\cmpeak \mid \ctpeak)$, constructed by passing each simulated image through the LSXPS count-rate determination tools and recording the measured counts $\cmpeak$. Each panel corresponds to a different true injected intensity $\ctpeak$, showing how the distribution of observed counts varies with source brightness. The red vertical line marks the corresponding injected true count value, $C_{T,\mathrm{peak}}$.}
    \label{fig:pmt}
\end{figure}

Each simulated image was passed to the LSXPS source detection software. If this detected an object within 5 times of the radial error radius of the simulated source and with a detection flag of \emph{Good} or \emph{Reasonable}\footnote{The criteria used by LSXPS to identify possible transients} then the count-rate measured by the LSXPS code ($I_M$) was recorded. This was converted to integrated counts ($C_M$) by multiplying by the image exposure time, and for each simulated $T$ the probability distribution $P(C_M \mid C_T)$ was constructed. Since the simulations are being used to represent the interval over which the LSXPS peak intensity was measured, these are the peak-snapshot quantities, i.e.\ $P(\cmpeak \mid \ctpeak)$. Note that this step involved binning the data since $I_M$ (and hence $C_M$) is background-subtracted and corrected for instrumental effects, and thus continuously distributed; we chose bin widths of one count. Examples of the distributions thus constructed are shown in Fig.~\ref{fig:pmt}.

\subsection{Inferring the true source intensity.}
\label{sec:posterior}
We can use our simulations to infer the true brightness of a source given its measured value, using Bayes’ theorem:
\begin{equation}
P(T \mid M) = \frac{\mathcal{P}(T)\, P(M \mid T)}{P(M)}
\label{eqn:trueintensity}
\end{equation}
\noindent
where $T$ and $M$ are the true and measured source brightness, $P(M \mid T)$ is the probability distribution derived from our simulations, $\mathcal{P}(T)$ is the Bayesian prior, and $P(M)$ is a normalisation factor ensuring that the total probability integrates to unity.  

The prior $\mathcal{P}(T)$ is the relative probability that a source of true brightness $T$ exists, reflecting the fact that faint objects are more numerous than bright ones. We calculate this using the empirical two-break $\log N - \log S$ relation from \cite{Mateos2008}. Our goal is to measure the probability distribution of the peak brightness, and as noted above, we consider this in terms of integrated counts. Thus, Equation~\ref{eqn:trueintensity} becomes:

\begin{equation}
P(\ctpeak \mid \cmpeak) = \frac{\mathcal{P}(\ctpeak)\, P(\cmpeak \mid \ctpeak)}{P(\cmpeak)}.
\label{eqn:finalP}
\end{equation}

and we calculate this for each simulated $T$ value, $T_0, T_1,\ldots T_{84}$

To determine $\mathcal{P}(\ctpeak)$ for $T_\mathrm{peak}=T_x$ we integrate the $\log N - \log S$ curve over half a bin either side\footnote{For $T_0$ we create a symmetric interval; i.e.\ the lower bound of the integral is $\log T_0 - (\log T_1 - \log T_0)$, and by analogy for $T_{84}$.} of $T_x$, that is, where:
\begin{equation}
    \frac{\log T_{x-1} + \log T_x}{2} \le \log S \le \frac{\log T_{x} + \log T_{x+1}}{2}
    \label{eqn:binrange}
\end{equation}
and divide this by the $\sum_x  P(C_{T_x})$. Note that the $\log N - \log S$ is in flux units, therefore we convert our measurements to flux using a typical AGN spectrum: an absorbed power-law with $N_H=3\tim{20}$ cm$^{-2}$, $\Gamma=1.7$ \citep{Evans2014}.

$P(\cmpeak \mid C_{T_x})$, the probability that we would measure \cmpeak\ photons if the source's true (peak) brightness were $T_x$, can simply be read from the simulations shown in Fig.~\ref{fig:pmt}. Note, however, that these results have to be binned (since it is the net photon count, which has a continuous distribution, that is measured); we accept simulations with $C_M = \cmpeak \pm1$. The normalisation, $P(\cmpeak)$, is simply $\sum_x{\mathcal{P}(C_{T_x})\, P(\cmpeak \mid C_{T_x})}$. Note also that the total probability simulated is $\sum_{{T_{\rm min}}}^{T_{\rm max}}{P_{\rm pois}(M|T)}$, so we also renormalise by this. We can thus determine the posterior distribution of $P(\ctpeak \mid \cmpeak)$ over our range of simulated $T$. 

For a real transient with measured intensity $\impeak$ and historical upper limit $L$,  we set $\ctpeak = \itpeak E$, where $E$ is the exposure time corresponding to the measurement of $\itpeak$. The posterior distribution is then computed as described above, and the integrated probability for $\ctpeak > LE$ is obtained. Since $LE$ rarely coincides exactly with a discrete $T$ value, we interpolate linearly between the adjacent bins on either side of $T = LE$ to estimate the fractional probability in the boundary bin. The resulting quantity, $P(\ctpeak > LE \mid \cmpeak)$, represents the probability that the true source intensity exceeds the historical ($3\sigma$) upper limit. For convenience we refer to this hereafter as \pnew.

\section{Results}
\label{sec:results}

\begin{figure*}
    \centering
    \includegraphics[width=0.45\linewidth]{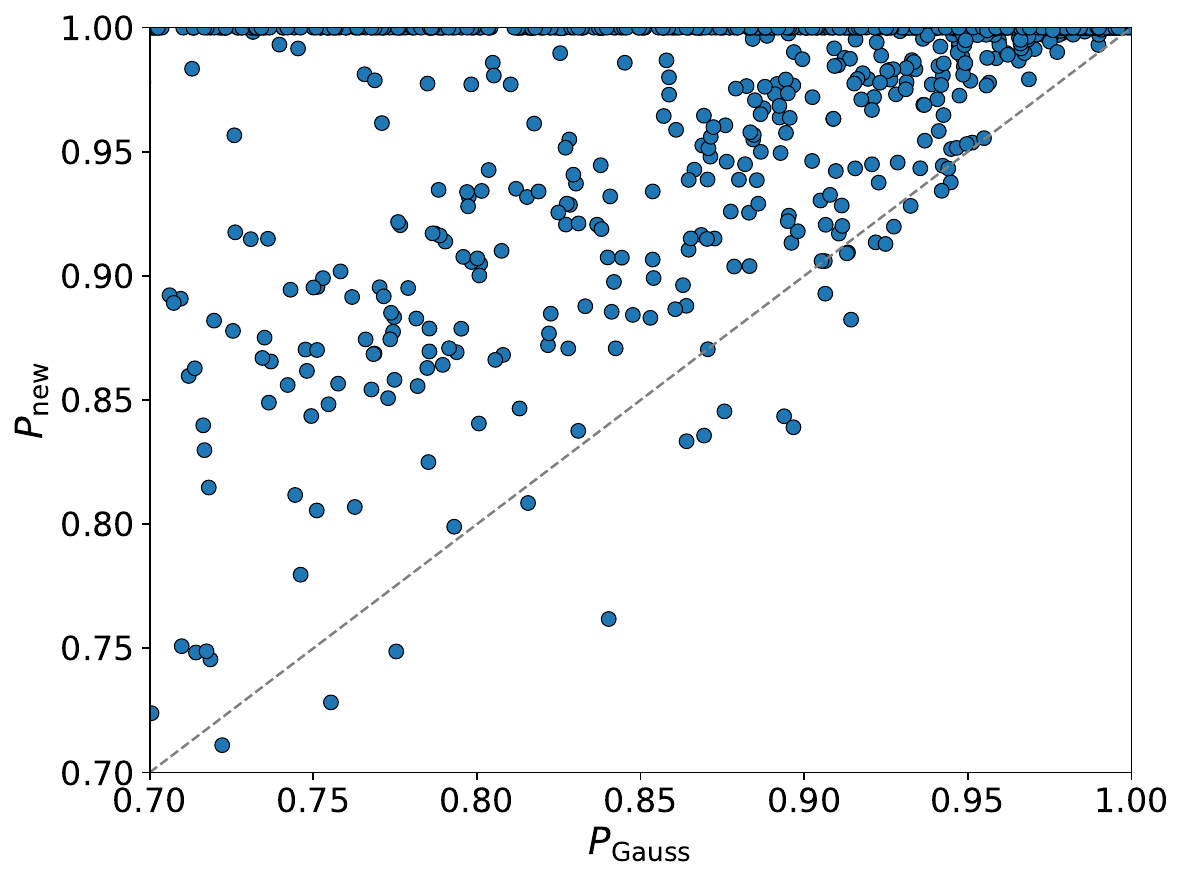}
     \includegraphics[width=0.45\linewidth]{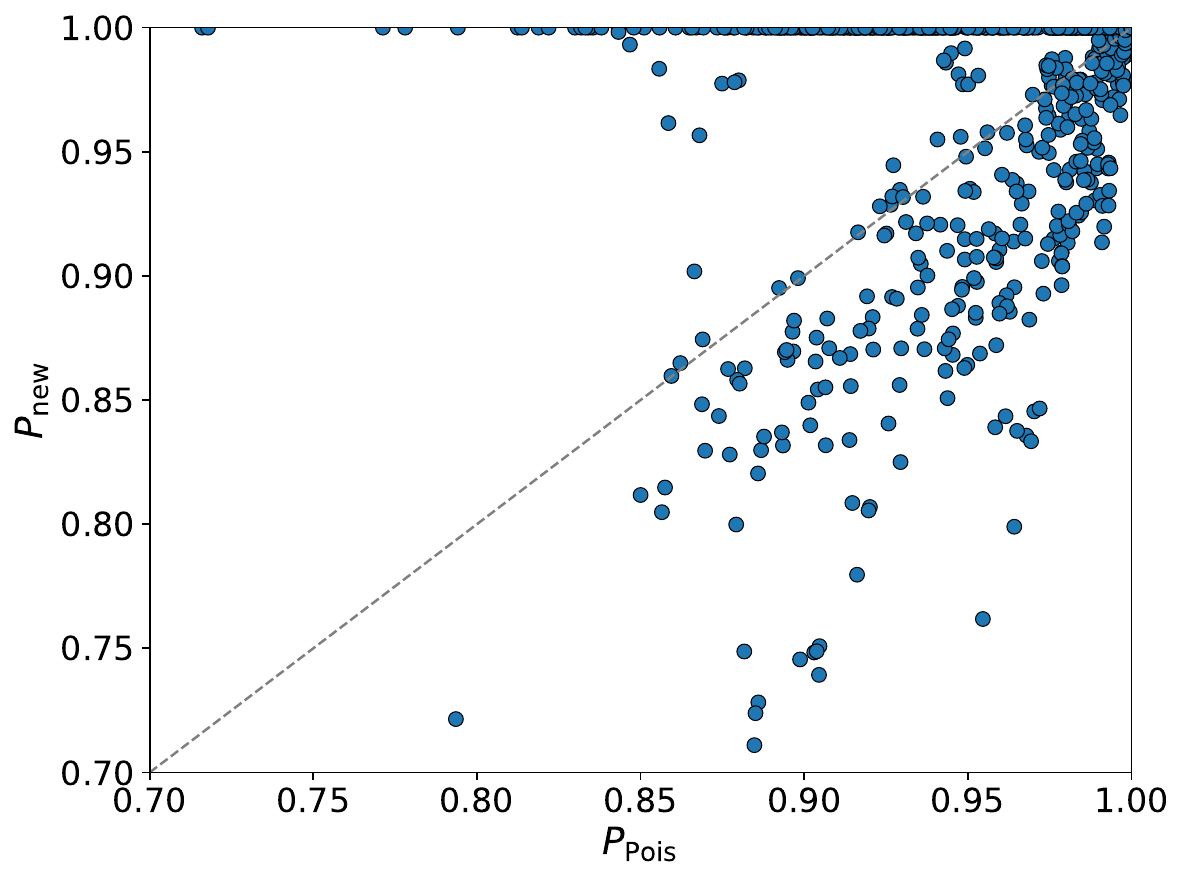}
     \caption{The probability that a low-significance transient candidate
really is brighter than the historical 3-$\sigma$ upper limit $[P(T>L)]$, calculated using our simulation-based approach ($P_{\mathrm{new}}$), against the value derived assuming the measured count-rate uncertainties are \emph{(Left)}: Gaussian and \emph{(Right)}: Poisson.}
    \label{fig:prob_vs_gauss}
\end{figure*}

\begin{figure}
    \centering
    \includegraphics[width=\linewidth]{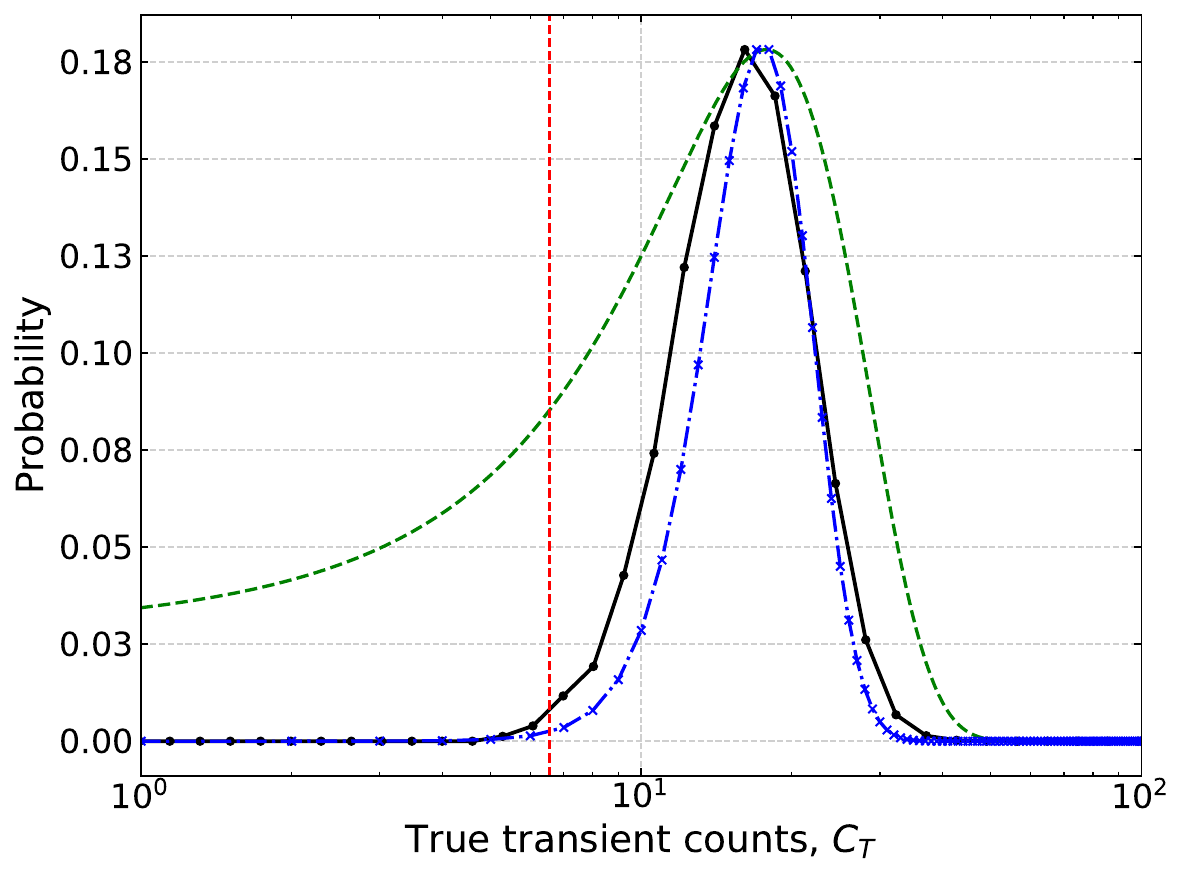}
    \caption{Comparison of the probability distributions for a representative low-significance transient candidate, showing the impact of correcting for Eddington bias. The black curve shows our bias-corrected probability distribution $P(C_{\mathrm{T}} \mid C_{\mathrm{M}})$, while the blue dash-dotted and green dashed lines show the corresponding Poisson and Gaussian distributions, respectively, assuming a mean of $C_{\mathrm{M}} = 18$ counts and Gaussian $\sigma = \sqrt{C_{\mathrm{M}}}$. The red dashed vertical line marks the historical $3\sigma$ upper limit. The Gaussian approximation clearly overestimates the probability at low true counts, while the Poisson form better captures the low-count regime and is therefore closer in shape to the corrected distribution. However, it does not include the source-count prior or selection effects, and so still remains biased. Our corrected distribution shifts slightly toward lower $C_{\mathrm{T}}$ values, reflecting the removal of the Eddington bias and yielding a lower overall probability above the upper limit.}

    \label{fig:fit}
\end{figure}

We applied our Bayesian framework to 1{,}757 transient candidates detected by LSXPS between 2022 April 1 and 2025 December 16, calculating $P(T>L)$ for each. Of these, 955 (54\%) are classed as low significance. Na\"\i vely, we expect this to yield lower probabilities than the current LSXPS system does since we have corrected for the Eddington bias, which artificially increases the measured flux and hence significance of sources near the detection threshold. The live LSXPS system does not report probabilities, but rather `outburst significance': {$\frac{\impeak - L}{\sigma_{\impeak}}$}, reported in the form $n \sigma$ which the LSXPS team, who categorizes the transients, tends to interpret as a Gaussian significance. In Fig.~\ref{fig:prob_vs_gauss} we show our probability \pnew\ against those derived using a Gaussian assumption. The dramatic increase in significance from our approach arises because, the majority of the transient candidates are in the Poisson, not the Gaussian regime \footnote{Note that LSXPS does not claim that the errors are Gaussian; however, reporting significances in terms of $\sigma$ does, with hindsight, make this assumption inevitable.} and the latter has an unphysically broad low-count tail (Fig. \ref{fig:fit}). It is, therefore, more meaningful to compare our Bayesian results with the current LSXPS approach using Poisson statistics for the latter, so we  calculated $P_{\rm pois}(\ctpeak>L)$ for each transient with mean $\ctpeak$; the result is shown in Fig.~\ref{fig:prob_vs_gauss} (right panel).

Overall, 44.3\% of the low significance transient candidates show a decrease in inferred significance relative to the original LSXPS values. Importantly, these revised significances are now more reliable, as the biases that previously inflated the measurements have been explicitly accounted for. A substantial subset of these sources are driven to zero probability: in total, 119 (12.5 \%) candidates have $P_{\mathrm{new}} = 0$, corresponding to all sources whose inferred significance falls below the 0.683 threshold once the bias correction is applied. Surprisingly, the remaining candidates show an increase in significance. We return to this behaviour in Section~\ref{sec:discussion}, and discuss the origin of both the zero-probability sources and the increased-significance cases in detail in Section~\ref{sec:Det}.

\begin{figure}
    \centering
    \includegraphics[width=\linewidth]{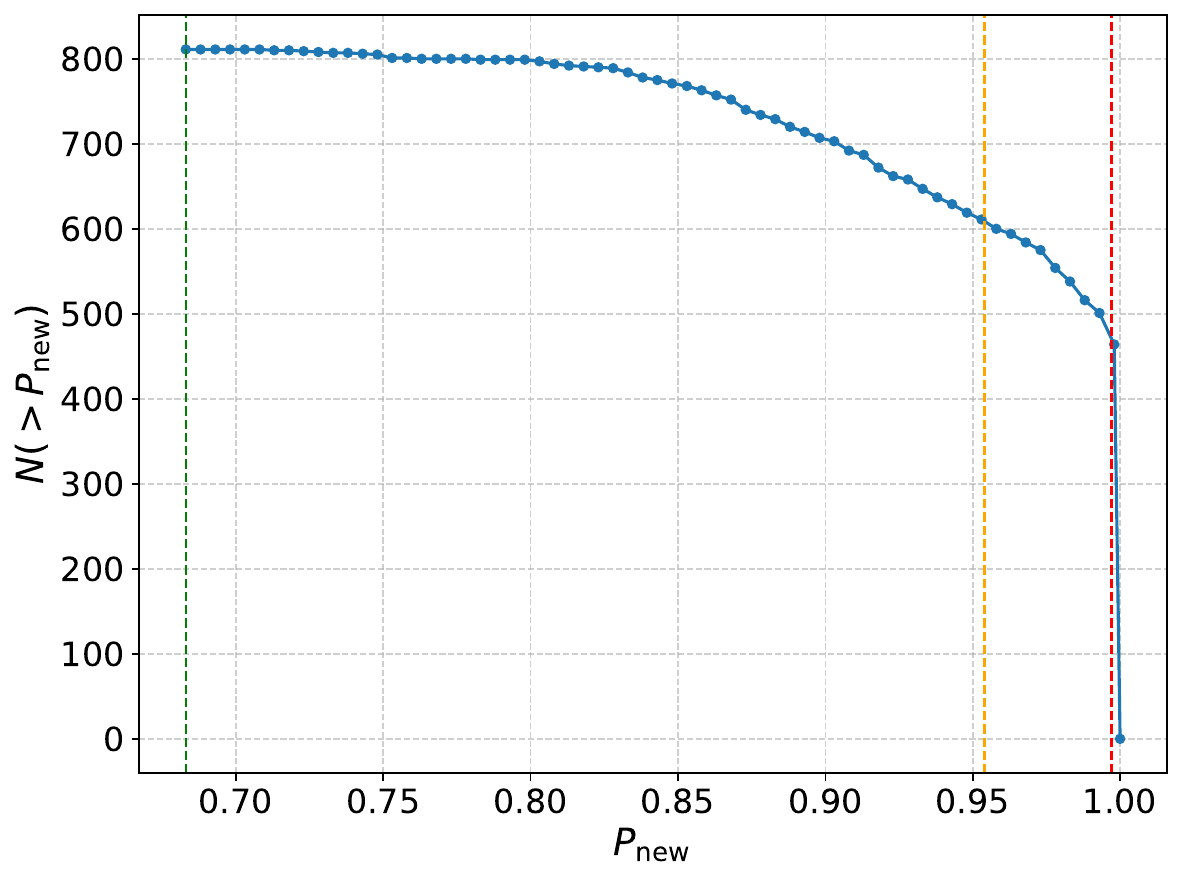}
\caption{The number of LSXPS low-significance transients which our system identifies as transient, as a function of the probability assigned by our system}, $P_{\mathrm{new}}$. Vertical lines indicate standard Gaussian thresholds for 1$\sigma$ (68.3\%), 2$\sigma$ (95.4\%), and 3$\sigma$ (99.7\%) confidence levels.
    \label{fig:recoveredtransients}
\end{figure}

The meaning of $P(T>L)$ merits some consideration, as it does not give the probability that a given source is transient and, interpreted as such, will likely underestimate that probability. First, note that the definition of a source as `transient' (at any wavelength) is a somewhat qualitative statement. Generally, what we mean is that it is newly-detected at a location where previously there had been no detection, down to a (significantly) deeper flux than its current appearance. However, this definition encompasses both sources which were previously `on' but too faint to detect with the equipment that has previously observed their location, and sources which were previously emitting no radiation. In reality most `transients' probably do emit pre-outburst \footnote{Few people would argue that a supernova is not a transient, for example, but their progenitors, while undetected  are still emitting, just at a level too faint for us to detect (until Betelgeuse or Eta Carina obliges).} so the question is one of \emph{how much} the flux has increased by -- which can only be stated probabilistically since by definition we have no historical measurement, only a limit. Second, while the upper limits we use are all reported `at the 3-$\sigma$ level', that is, they are given with 99.7\%\ confidence, we cannot make reliable assumptions about the underlying distribution of this probability and therefore cannot combine it with our $P(T>L)$ values to get some $P(T>\rm{quiescent})$ estimate\footnote{i.e.\ $P(T>L)=0.954$ does not correspond to a 2-$\sigma$, (2+3)-$\sigma$ or $(\sqrt{2^2+3^2})$-$\sigma$, probability that the source is a transient.}. Rather, $P(T>L)$ is literally the probability that the source is brighter than $L$, which happens to be a  99.7\%\ confidence upper limit; it is for the user to decide the threshold at which they wish to treat a candidate as a transient. Fig.~\ref{fig:recoveredtransients}  shows, for the transients classified as `low significance' in LSXPS but with $P(T>L) > 0.683$ (i.e. `$1-\sigma$`) using our approach, the number of transients as a function of $P(T>L)$. Even using a conservative setting (that a transient must be above the historical 99.7\%\ confidence upper limit with 99.7\%\ confidence), we find that 500 out of 955 low significance transients would be newly selected. This corresponds to an \emph{approximate eight-fold increase} over the 64 previously confirmed transients.

\section{Discussion} \label{sec:discussion}
Of the 955 `low-significance' transient candidates in our sample, 55.7\% increase in significance using our method compared to the simple Poisson test; the others decrease. This is the result of two competing effects. On the one hand, our prior [$\mathcal{P}(\ctpeak)$ in Equation~\ref{eqn:finalP}] accounts for the Eddington Bias, i.e.\ that fainter sources which are `upscattered' in measured counts due to Poisson processes are more populous than the brighter sources which are downscattered. This prior extends the low-intensity tail of the intensity probability distribution. On the other hand, $P_{\rm pois}(M|\mu=T) < P_{\rm pois}(T|\mu=M) $ when $T<M$; and the probability of detection $P({\rm det}|M)$ is not a binary function. These effects, which are empirically captured by our simulations (as $P(\cmpeak | \ctpeak)$ in Equation~\ref{eqn:finalP}), curtail the low-intensity tail of the intensity probability distribution. Whether the significance of a given transient increases or decreases depends on which of these effects dominates.

\subsection{Dependence on Simulation and Background Properties}
\label{sec:simdep}

Our initial set of simulations made use of a single seed XRT image, with an exposure time of 2016 s; we recorded the results from this in terms of integrated counts (rather than count-rates), allowing us to apply these results to any detected transient candidate. We do not expect the choice of seed image to significantly affect our results for two reasons. First, the source-detection performance of LSXPS is photon-limited out to exposure times
$\sim10^5$~s (see fig.~7 of \citealt{Evans2020}); thus we do not expect the probability of detection as function of the number of source photons to change with exposure. Second, the $C_M$ values are background subtracted, so the variation in background counts that is expected with changing exposure should not affect the measured value -- assuming that the background is accurately known.

\begin{figure*}
    \centering
    \includegraphics[width=0.85\linewidth]{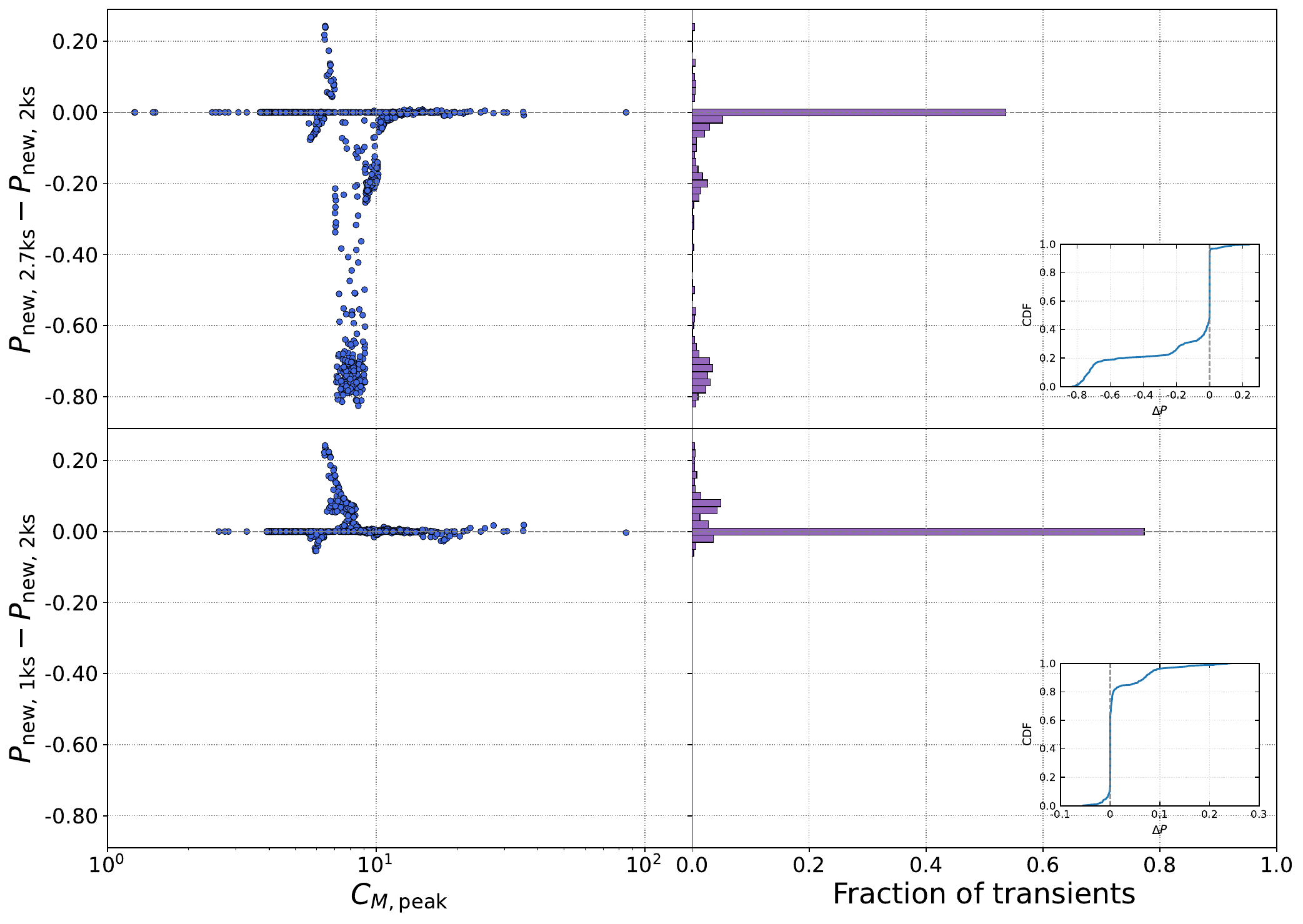}
   
    \caption{The effect of the seed-image exposure time on the inferred transient probabilities.
\emph{Left:} the difference in \pnew\ between simulations using alternative seed images and the original simulations, shown as a function of the measured peak counts, \cmpeak, for seed images with exposures of 2.7~ks (top) and 1~ks (bottom).
\emph{Right:}  Histograms of the corresponding probability differences.
\emph{Insets:} Cumulative distribution functions (CDFs) of $\Delta P$, illustrating the fraction
of sources whose inferred probabilities change by a given amount}

    \label{fig:Compare}
\end{figure*}

To determine whether this expectation is correct, we repeated our entire analysis twice using different seed images for the simulations, with exposure times of 1~ks and 2.7~ks. Figure~\ref{fig:Compare} shows the difference between \pnew\ ($\Delta P$) derived from these simulations and that obtained from our original analysis for the low-significance transients. For the majority of sources, the inferred probability is only weakly affected by the choice of seed image: for 65.9\% (89.7\%) of low-significance transients, $|\Delta P| < 0.1$ when using the 2.7~ks (1~ks) seed. This includes cases for which the original 2~ks analysis yielded $P_{\mathrm{new}} = 0$, but a non-zero probability is recovered when alternative seed images are used. Two notable features are, however, apparent: large excursions in $\Delta P$ at $C_{\mathrm{M,peak}} \lesssim 10$ counts, and a pronounced asymmetry in the distribution of probability differences.

\begin{figure*}
    \centering
    \includegraphics[width=0.32\linewidth]{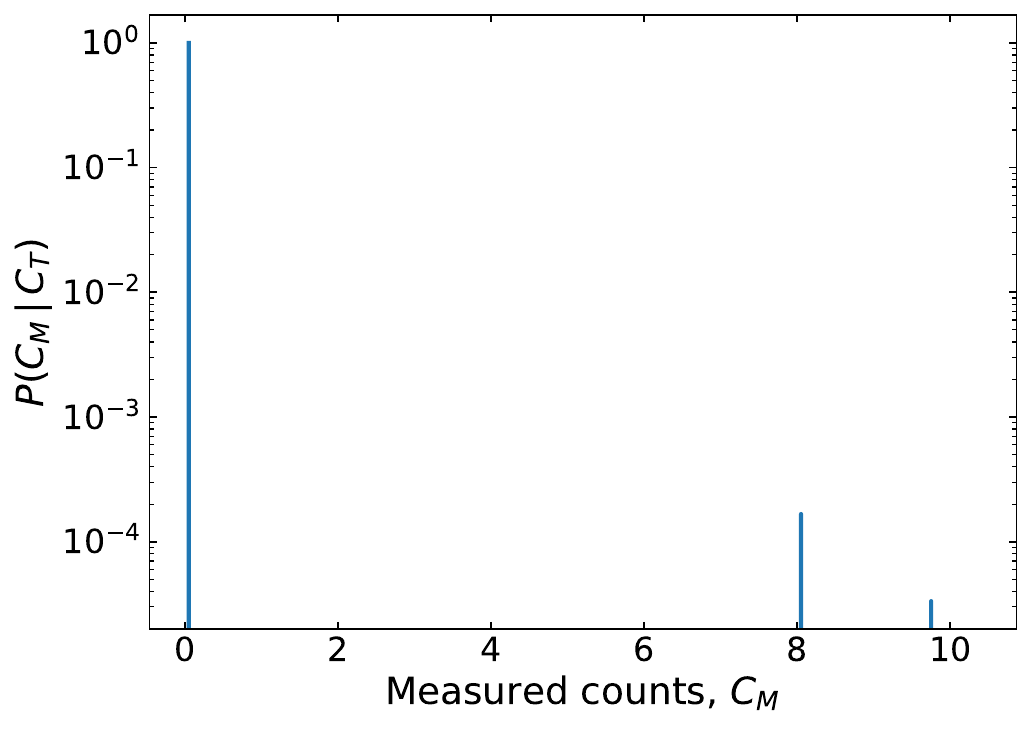}
    \includegraphics[width=0.32\linewidth]{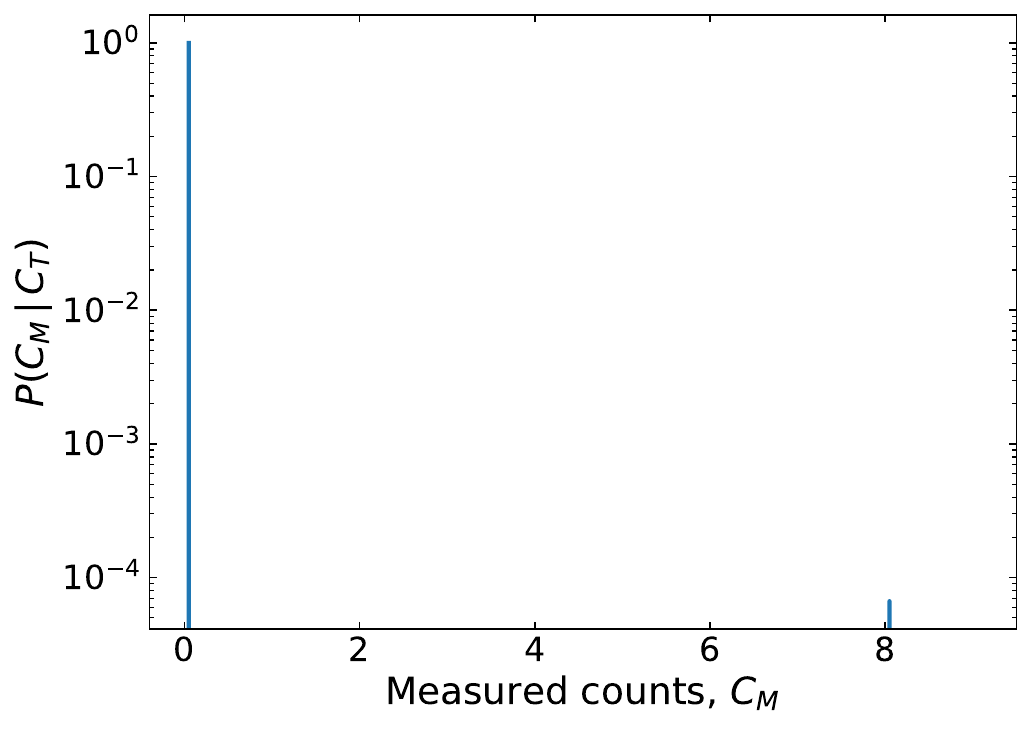}
    \includegraphics[width=0.32\linewidth]{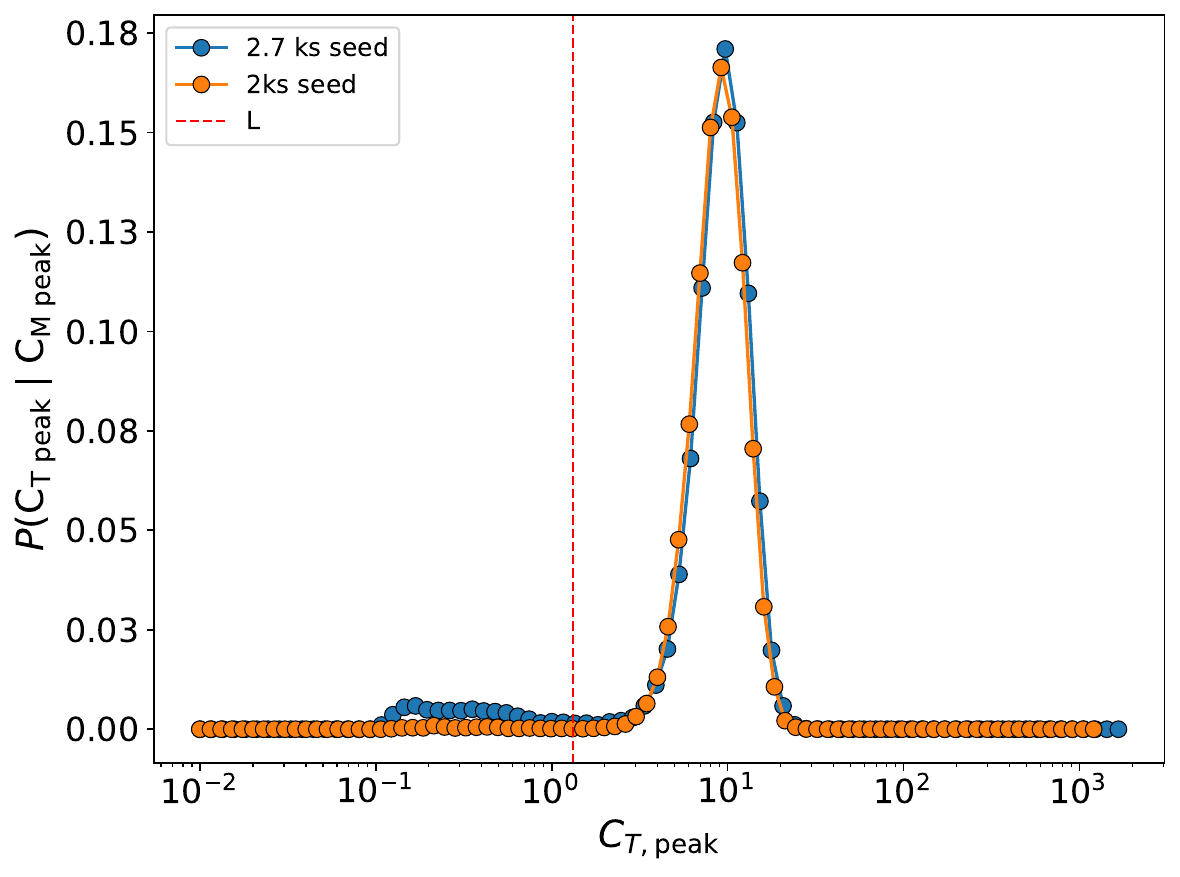}
    
\caption{Illustration of how near-miss sources and background
underestimation distort the measured--true count relationship, shown
here for simulations with $C_T \approx 1$.
\textit{Left:} examples of $P(C_{\mathrm{M,peak}} \mid C_{\mathrm{T,peak}})$ for simulations affected by a near-miss source, showing sharp spikes at $M \gg T$, where a faint undetected cluster artificially boosts the measured counts.
\textit{Middle:} an example of $P(C_{\mathrm{M,peak}} \mid C_{\mathrm{T,peak}})$ for a case affected by an underestimated background, again producing narrow probability spikes at measured counts far exceeding the true source counts.
\textit{Right:} the resulting inferred distribution $P(C_{\mathrm{T,peak}} \mid C_{\mathrm{M,peak}})$ for a representative transient, showing that the 2.7~ks simulation (blue) develops a faint excess toward lower true intensities compared to the 2~ks case (orange). This excess probability at low $T$ suppresses the inferred transient significance. The red dashed line marks the historical 3-$\sigma$ upper-limit threshold, $L$.}
\label{fig:P27/2}
\end{figure*}

The first of these effects arises from a small subset of simulations in which few counts are simulated but many are measured. In such cases, $P(\cmpeak \mid \ctpeak)$ acquires non-negligible probability at values of \cmpeak\ far higher than expected for a Poisson distribution with mean \ctpeak. When inverted, this leads to a posterior $P(\ctpeak \mid \cmpeak)$ with excess weight at low \ctpeak, an effect further amplified by the prior $\mathcal{P}(\ctpeak)$. In all such instances, the simulated source lies close to one of a small number of specific locations in the seed images. Inspection of these regions reveals clusters of 4--6 photons, consistent with faint sources just below the LSXPS detection threshold. The addition of only 1--2 photons from a simulated transient within a few arcseconds of such a cluster can therefore trigger a detection with a measured intensity of 5--8 counts. An example of this behaviour is shown in Fig.~\ref{fig:P27/2}.

Both these near-miss sources and small inaccuracies in the background estimate can produce unrealistically high measured counts for a given true source intensity, giving rise to sharp spikes in $P(M \mid T)$ at $M \gg T$. Figure~\ref{fig:P27/2} illustrates both effects. The left and middle panels show representative examples of $P(\cmpeak \mid \ctpeak)$, the first dominated by a near-miss source adjacent to an undetected cluster, and the second by an underestimated background level. In both cases, narrow peaks appear at large $M$. The right-hand panel shows the corresponding inferred $P(\ctpeak \mid \cmpeak)$ for a representative transient, where the 2.7~ks simulation exhibits a faint excess toward lower true intensities that is absent in the 2~ks result, reflecting the combined impact of these effects on the inferred transient significance.

\begin{figure*}
    \centering
    \includegraphics[width=0.9\linewidth]{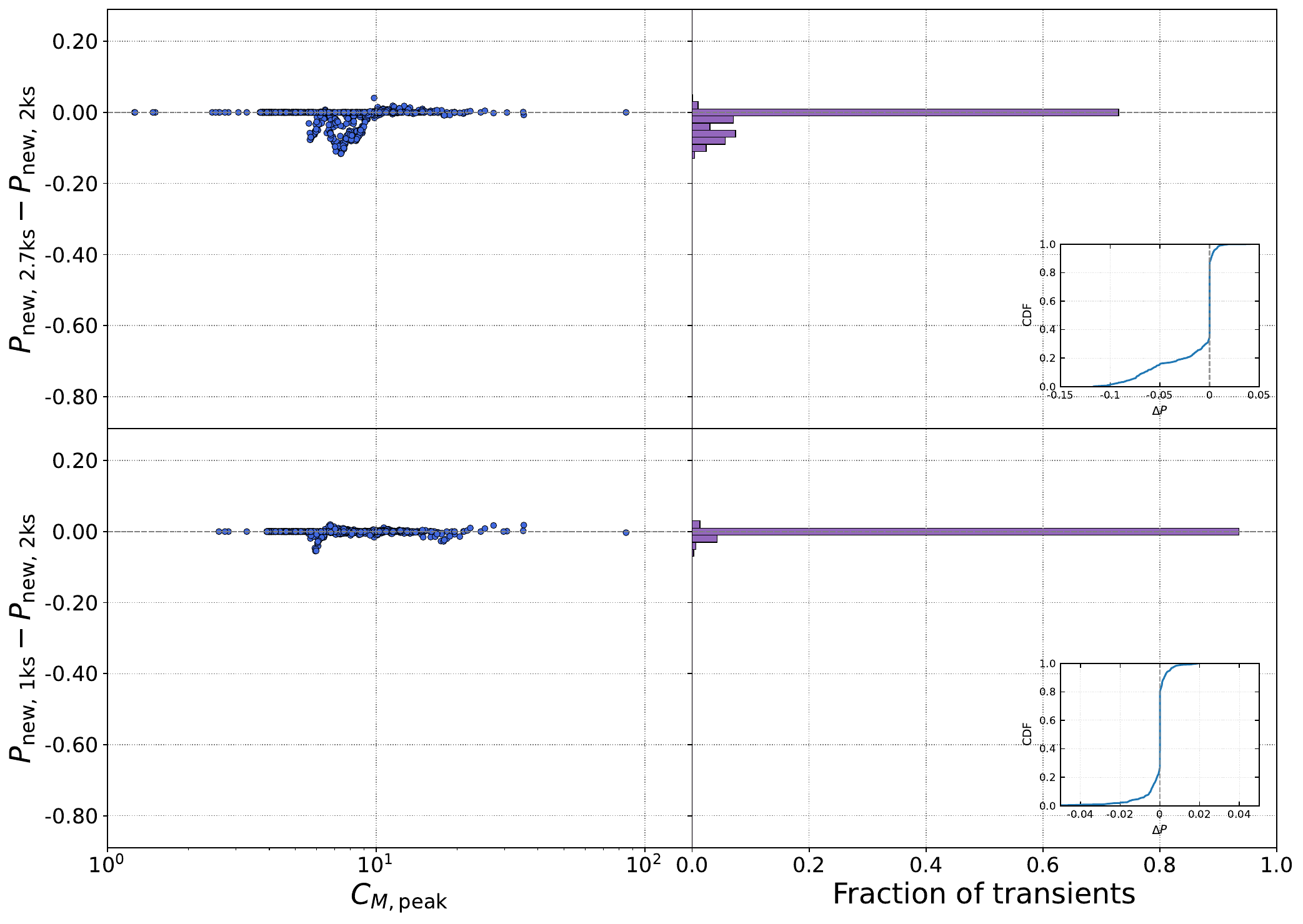}
\caption{Effect of removing near-miss sources on the inferred transient probabilities. 
\emph{Left:} the difference in $\pnew$ relative to the nominal 2~ks simulations,
$\Delta P$, as a function of measured peak counts $C_{\mathrm{M,peak}}$,
for simulations seeded with 2.7~ks (top) and 1~ks (bottom) images.
\emph{Right:} the corresponding distributions of $\Delta P$, with inset panels showing the cumulative distribution functions.
Only simulations in which injected sources coinciding with pre-existing, undetected
photon clusters have been excluded are shown. }

\label{fig:notail}
\end{figure*}

To verify that this `near-miss' condition is responsible for the large excursions seen
in Fig.~\ref{fig:Compare}, we excluded from our analysis those simulations in which an
injected source fell within $5$~arcsec of one of the identified pre-existing
photon clusters, and repeated the calculations.
The large spike is almost entirely removed (Fig.~\ref{fig:notail}), confirming that these
near-detections are indeed responsible for the extreme deviations. Once these cases are
excluded, 97.5\% (94.6\%) of the results from the 2.7~ks (1~ks) simulations differ by less
than 0.1 from the corresponding values obtained using the original 2~ks seed image.

This is a realistic scenario that can also occur in real XRT data: a faint transient or variable source may lie along the same line of sight as a steady source which is just below the LSXPS detection threshold, thus triggering the detection of an apparently single `transient' source, with a measured intensity much greater than that of the transient/variable; thus in principle there is no special need to remove these simulations from our calculations. However, we remove these near-miss regions in all subsequent simulations and figures for several reasons. First, because they would mask the impact of the other factors we discuss. Second,  in real applications there is often pre-existing \swift\ coverage, so such cases can be quickly identified by examining those previous. Finally, we do not know the occurrence rate of near-misses as a function of exposure time and field properties, so retaining them would introduce an uncontrolled, exposure-dependent contribution to the inferred probability distributions.\footnote{The LSXPS web pages for transients, will provide an option to view the `including near-miss' probability, but it will not be displayed by default.}

The second effect---the asymmetry in $\Delta P$ that remains even after removing near-miss cases (Fig.~\ref{fig:notail}) is most naturally explained by small systematic errors in the local background level. In LSXPS, the measured peak counts $C_{\mathrm{M,peak}}$ are obtained by summing photons within an intensity-dependent radius around the source position and subtracting the background estimate from the LSXPS background map evaluated over the same region (see \citealt{Evans2014}). If this background is underestimated, then $C_{\mathrm{M,peak}}$ is biased high (and vice versa), which propagates into the inferred transient probabilities. This is closely
related to the ``near-miss'' behaviour: in both cases, the simulations can yield a non-negligible probability of measuring $M\gg T$ in $P(M\mid T)$ (cf.\ Fig.~\ref{fig:P27/2}).

If the asymmetric $\Delta P$ distributions are driven by background errors, then seed images that tend to give lower \pnew\ should also show a tendency for the background map to underestimate the realised background. To test this, for every accepted detection in each simulation set (1, 2, and
2.7~ks seeds) we compared the background-map prediction to the counts actually present in the seed image, using exactly the same region over which LSXPS measures the source intensity. We define
\begin{equation}
\Delta B \;=\; C_{\rm bgmap} - C_{\rm seed},
\end{equation}
where $C_{\rm seed}$ is the number of counts measured from the seed image in the source region and $C_{\rm bgmap}$ is the background-map prediction integrated over the same region. Negative $\Delta B$ therefore means the map \emph{underestimates} the true background, so background subtraction yields an artificially large net source signal and hence an overestimated measured intensity; positive $\Delta B$ corresponds to
the opposite effect.

\begin{figure}
    \centering
    \includegraphics[width=\linewidth]{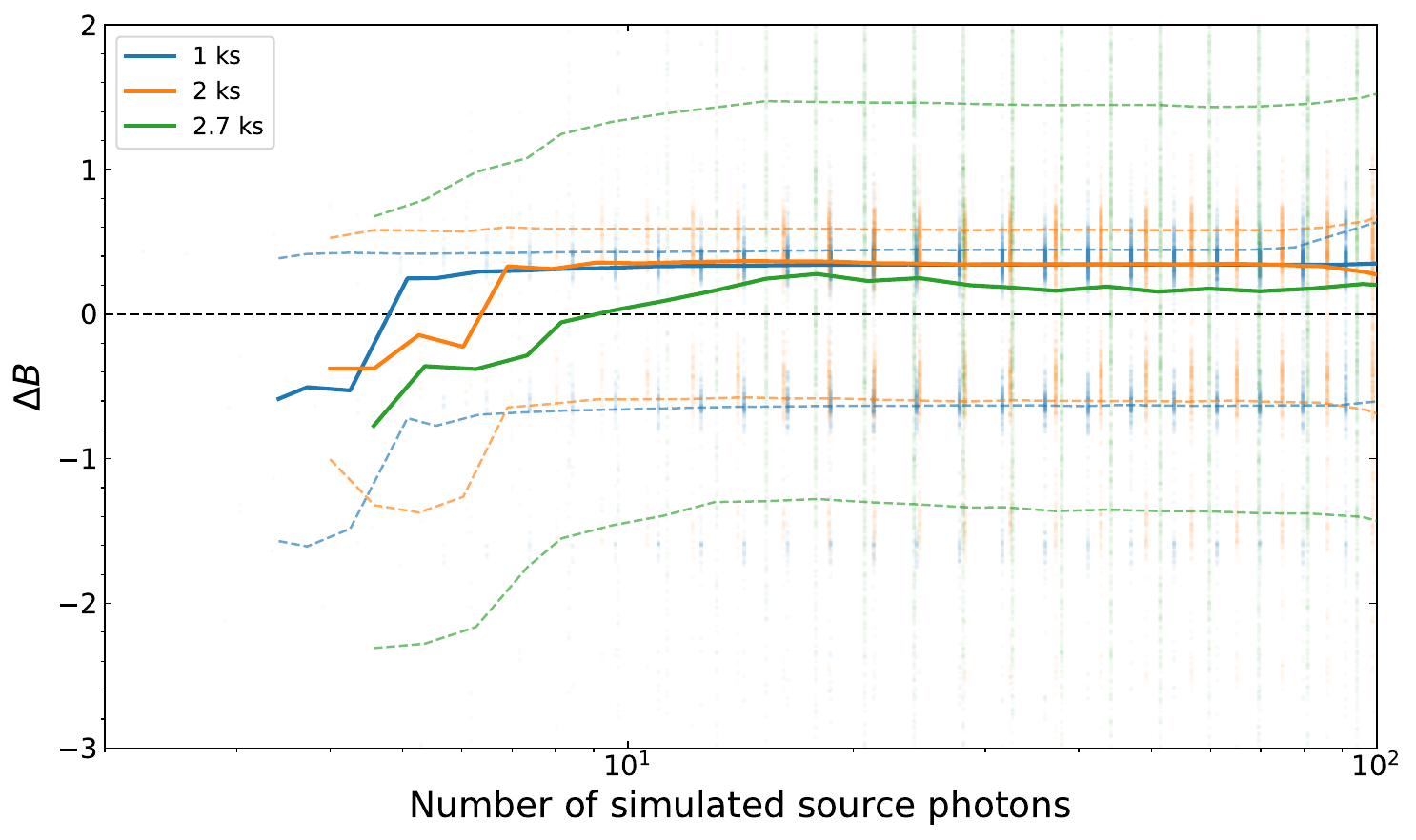}
    \caption{Background-map residuals as a function of simulated source strength.
Points show $\Delta B = C_{\rm bgmap}-C_{\rm seed}$, measured in the same
(intensity-dependent) region used by LSXPS to derive the source intensity, plotted against the number of injected source photons. Solid curves show the median $\Delta B$ for each seed exposure, and dashed curves indicate the
$\pm1\sigma$ scatter about the median.}

    \label{fig:bgplaceholder}
\end{figure}

The results are shown in Fig.~\ref{fig:bgplaceholder}, where $\Delta B$ is plotted against the true number of injected source photons. The median $\Delta B$ is close to zero for all seed exposures, but the distributions are visibly skewed, with negative tails whose extent depends on the seed image. The 2~ks seed is the tightest, whereas the 1~ks and especially the 2.7~ks seeds show broader, more extended negative excursions. This matters most for faint sources, where a mismatch of even a few background counts is comparable to the true source signal. Consistent with this, for $T\lesssim 5$ photons the 1~ks and 2.7~ks seeds more often yield $\Delta B<0$. Positive excursions ($\Delta B>0$) are also present, but are less extended in these simulations.

This behaviour provides a natural explanation for the exposure-dependent trends in \pnew. When $\Delta B<0$, the background is underestimated and the background-subtracted source counts, $C_{\mathrm{M}}$, are biased high. That increases the chance of obtaining an apparently large $C_{\mathrm{M}}$ from a modest true source intensity, which enhances the likelihood of $C_{\mathrm{M}}$ at fixed $T$ in the high-$C_{\mathrm{M}}$ tail. When this is mapped back to $P(T\mid C_{\mathrm{M}})$, and combined with the $\log N$--$\log S$ prior that favours faint sources, extra weight is assigned to lower true intensities and the integrated probability $P(T>L)$ is reduced. This explains why the 2.7~ks simulations, and to a lesser extent the 1~ks simulations, tend to give lower inferred transient probabilities than the 2~ks reference, despite similar median behaviour.

Having established that the differences between simulations arise primarily from small systematic offsets in the background maps, we conclude that the overall method is robust against variations in both exposure time and background estimation.

\subsection{Detection vs peak exposure time}\label{sec:Det}

In our simulations, the same dataset was used both to determine whether a source was detected, and to measure its peak count rate. This is not the case for LSXPS, in which source detection runs on as much of the full observation as has been downlinked (which has exposure \efull), whereas the peak count-rate is obtained either from this exposure or from any individual spacecraft orbit within it (exposure \epeak). For $\sim$75.4\%\ of the transient candidates in our low significance  sample (including all of those for which \pnew\ reported above was 0) $\efull > \epeak$. We cannot properly account for this in our simulations without knowing the true light curve of the source (since we simulate the true, not measured behaviour) or with some well-defined and justified Bayesian prior on the light curve shape, neither of which we have. Instead, we have explored two plausible cases to give an indication of the impact this has on the posterior significance.

\begin{itemize}
    \item \textbf{Case (a):} Assume that the source was active only during the interval over which the peak intensity was measured.
   That is, $\ctpeak = \ctfull$ and thus we need only consider the interval \epeak. This is identical to the baseline case already simulated.
   
    \item \textbf{Case (b):} Assume that the observed light curve exactly matches the true one, i.e.\ assume:  
    \begin{eqnarray}
    \frac{T_{\mathrm{peak}}}{T_{\mathrm{full}}} &=& \frac{M_{\mathrm{peak}}}{M_{\mathrm{full}}}. \nonumber \\
    T_{\mathrm{peak}} &=& T_{\mathrm{full}} \frac{M_{\mathrm{peak}}}{M_{\mathrm{full}}}
    \label{eqn:lcCorr}
    \end{eqnarray}
         
    Operationally, we replaced $\ctpeak$ and $\cmpeak$ in Equation~\ref{eqn:finalP} with $\ctfull$ and $C_{\mathrm{M,full}}$, ensuring that the probability of detection reflects the exposure in which the source was detected. The resulting distribution $P(\ctpeak)$ was then obtained by rescaling the $x$-axis of $P(\ctfull)$ using Equation~\ref{eqn:lcCorr}, after which we recomputed the probability that the true source intensity exceeded the historical upper limit.

\end{itemize}

\begin{figure}
    \centering
    \includegraphics[width=\linewidth]{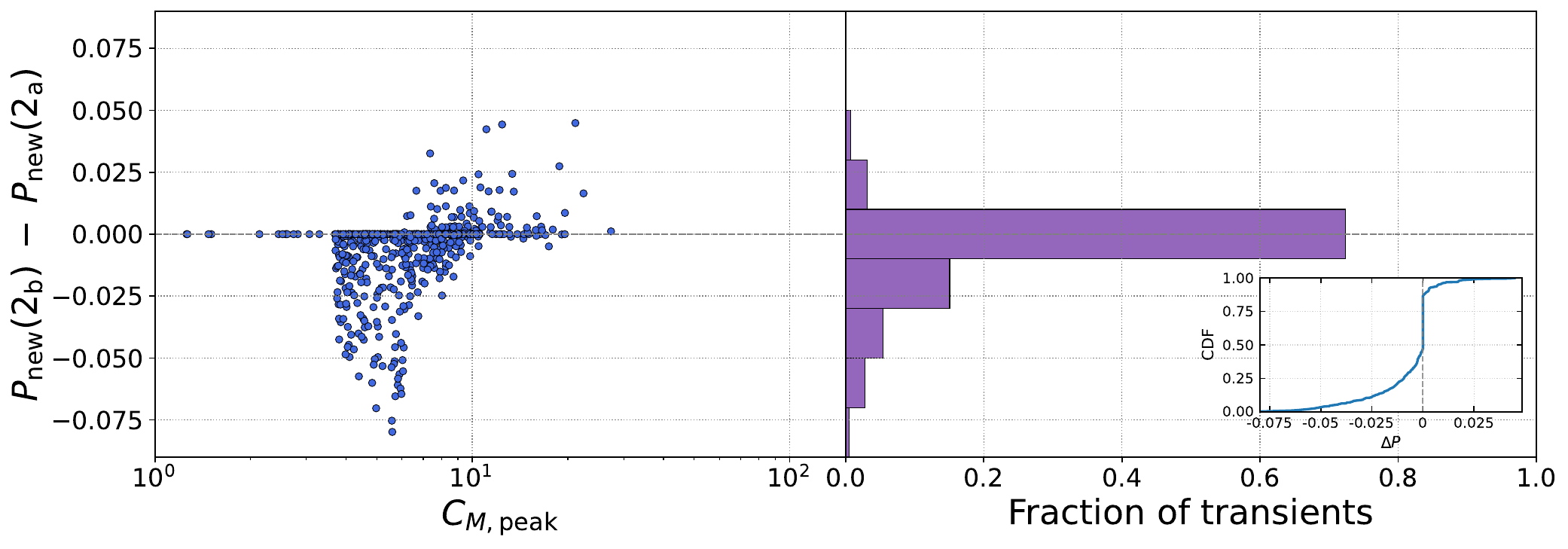}
    \caption{Comparison of inferred transient probabilities under different assumptions about source variability for sources with $\epeak<\efull$. The panel shows the difference between Case~(b) (measured using the full detection exposure) and Case~(a) (measured using the peak snapshot)
    Left-hand plots show the difference in probability as a function of measured counts, and right-hand panels show the corresponding distribution and cumulative fraction of sources.}
    
    \label{fig:case2}
\end{figure}

Figure~\ref{fig:case2} shows how the inferred transient probability changes in
Case~(b) relative to the nominal simulations (Case~a) for candidates with $E_{\mathrm{full}} > E_{\mathrm{peak}}$. For the majority of transients ($\sim70\%$), the change in \pnew\ is negligible ($\Delta P \simeq 0$). The remainder show a clear asymmetry: at low measured counts ($\cmpeak \lesssim 10$) Case~(b) typically yields lower probabilities than Case~(a), whereas above this regime the trend reverses and Case~(b) yields slightly higher \pnew.\footnote{Note that this plot does not include those
transients for which the nominal case gave $\pnew=0$; we return to those
shortly.}

\begin{figure*}
    \centering
    \includegraphics[width=0.45\linewidth]{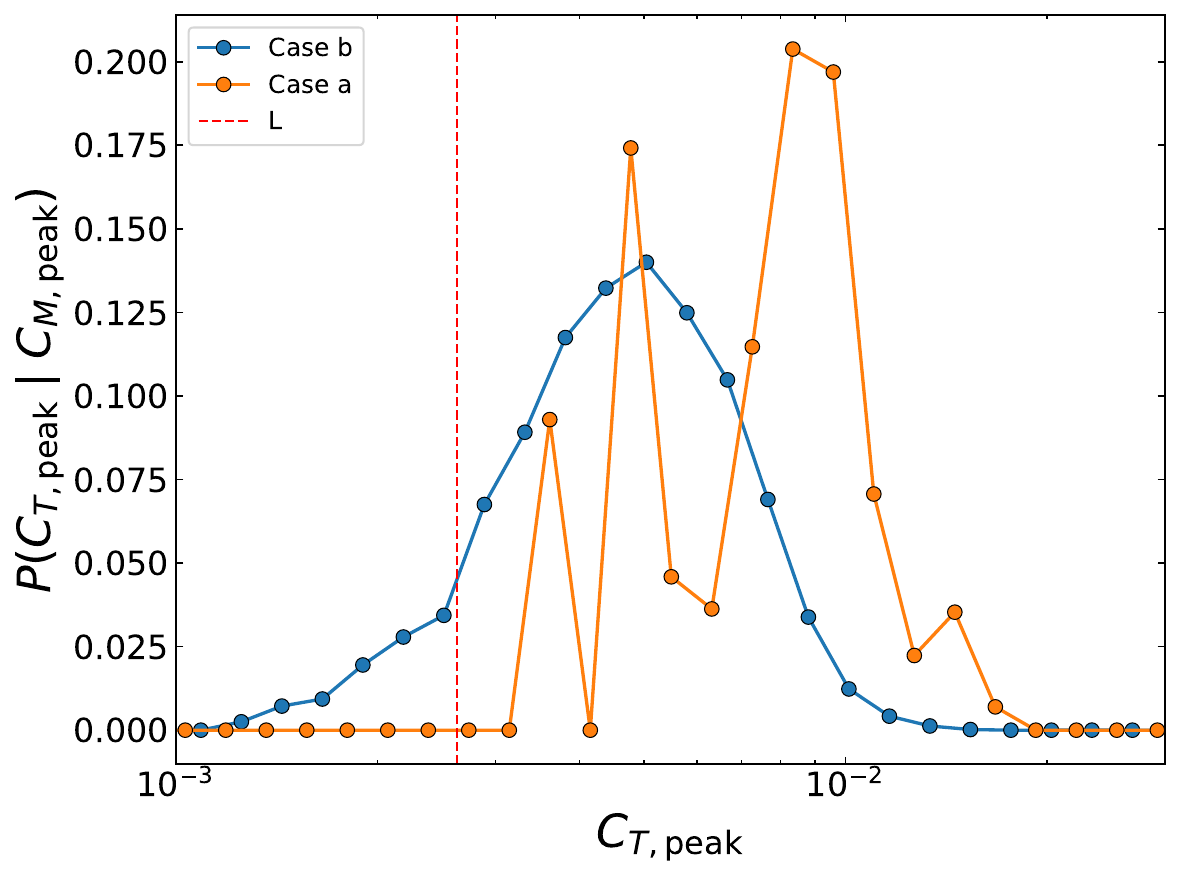}
    \includegraphics[width=0.45\linewidth]{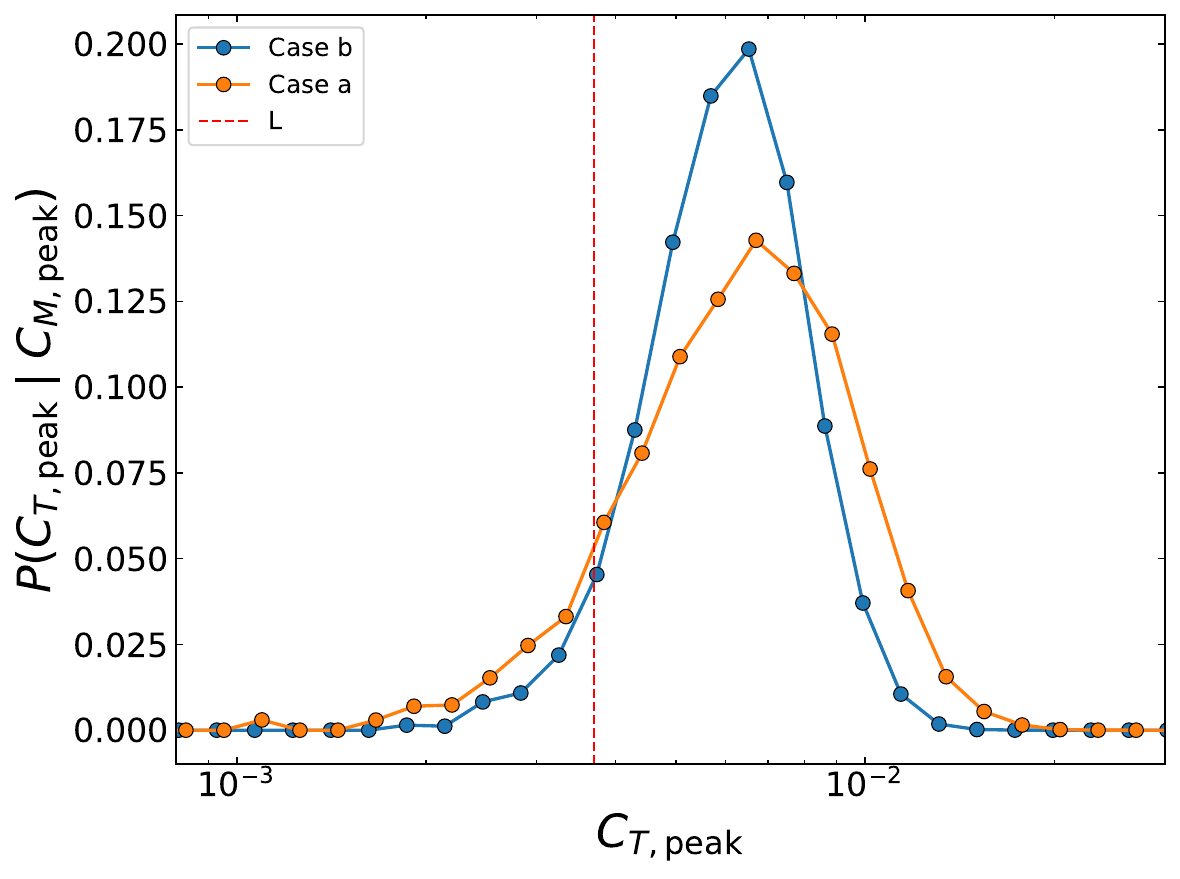}
   \caption{Example distributions illustrating why Case~(b) can either decrease or increase the inferred transient significance. We show the inferred $P(C_{\mathrm{T,peak}}\mid C_{\mathrm{M,peak}})$ for two representative candidates, comparing Case~(a) (orange; using the peak-snapshot exposure) and Case~(b) (blue; conditioning on the full detection exposure and mapping back to the peak). \textit{Left:} a faint candidate for which $P_{\mathrm{b}}<P_{\mathrm{a}}$; near the detection limit the Case~(a) curve is irregular because detections are sparse in the $(C_{\mathrm{T,peak}}<C_{\mathrm{M,peak}})$ regime, whereas Case~(b) is smoother and assigns more weight to lower true intensities, reducing the probability above the limit. \textit{Right:} a brighter candidate for which $P_{\mathrm{b}}>P_{\mathrm{a}}$; the rescaled Case~(b) distribution is slightly narrower than the Case~(a) distribution, giving a modest increase in the probability above the limit. The red dashed line marks the historical upper-limit threshold $L$.}

    \label{fig:pabrate}
\end{figure*}

This is because, for sources with $\cmpeak\lesssim10$, the probability of detection by LSXPS drops, so for simulations with $\ctpeak < \cmpeak$ the probability of detecting the source (and measuring \cmpeak\ counts) becomes both badly sampled and small, hence the posterior probability $P(\ctfull<\cmfull)$ is small. This is accurate -- however, the actual transient candidates were not \emph{detected} in this regime! They were detected with \cmfull\ (which is $>\cmpeak$) counts measured over \efull. Our case~(b) simulations use these larger values, and the parameter space $P(\cmfull | \ctfull < \cmfull)$ is fully sampled and can have significant non-negligible probability as shown in Fig. \ref{fig:pabrate}; as a result, $P(\ctfull<\cmfull)$ is higher than in case a and hence $P(\ctpeak > L)$ is lower.

For brighter transients, the $\ctpeak < \cmpeak$ space is fully sampled and the above effect is negated. In fact, some (10\%) of transients show $P_{\mathrm{b}} > P_{\mathrm{a}}$. This arises because in case b we must calculate the posterior $P(\ctfull | \cmfull)$ and then multiply the $x-$axis of the $(\ctfull, P(\ctfull | \cmfull)$ data by $\frac{\cmpeak}{\cmfull}$ to calculate $P(\ctpeak > L)$\footnote{Equivalently (and easier to visualise), we calculate $P(\ctfull > L \frac{\cmfull}{\cmpeak} | \cmfull)$.} However, the Poisson distribution (from which our posterior is derived) gets proportionally narrower for higher mean values (that is $l/\mu$ increases for increasing $\mu$, where $l$ is some arbitrary value on the cumulative distribution function which occurs before the peak). As a result, our rescaled posterior distribution is slightly narrower that measured natively for $\cmpeak$ (in the nominal case/case a), resulting in a small increase in $P(T > L)$.

It is also worth noting that Case~(b) also restores the candidates for which Case~(a) returned $\pnew=0$. These are objects where the transient was detected in the full downlinked dataset, but the peak snapshot alone sits so close to the detection threshold that our Case~(a) simulations contain essentially no realisations in which the source is both detected and yields the observed $\cmpeak$ counts. In practice this means the Case~(a) detection term drives $\pnew$ to zero. In Case~(b) we instead condition on the same $(C_{\rm M,full},E_{\rm full})$ measurement that actually triggered the LSXPS detection, so the relevant part of $P(\cmfull\mid \ctfull)$ is well sampled and the detection probability is non-negligible. As a result, all of the Case~(a) ``zero-probability'' candidates are recovered in Case~(b); in our sample the smallest value is $P_{\rm new,b}\simeq 0.80$.

\subsection{Overall robustness of the transient probabilities}

\begin{figure}
    \centering
    \includegraphics[width=\linewidth]{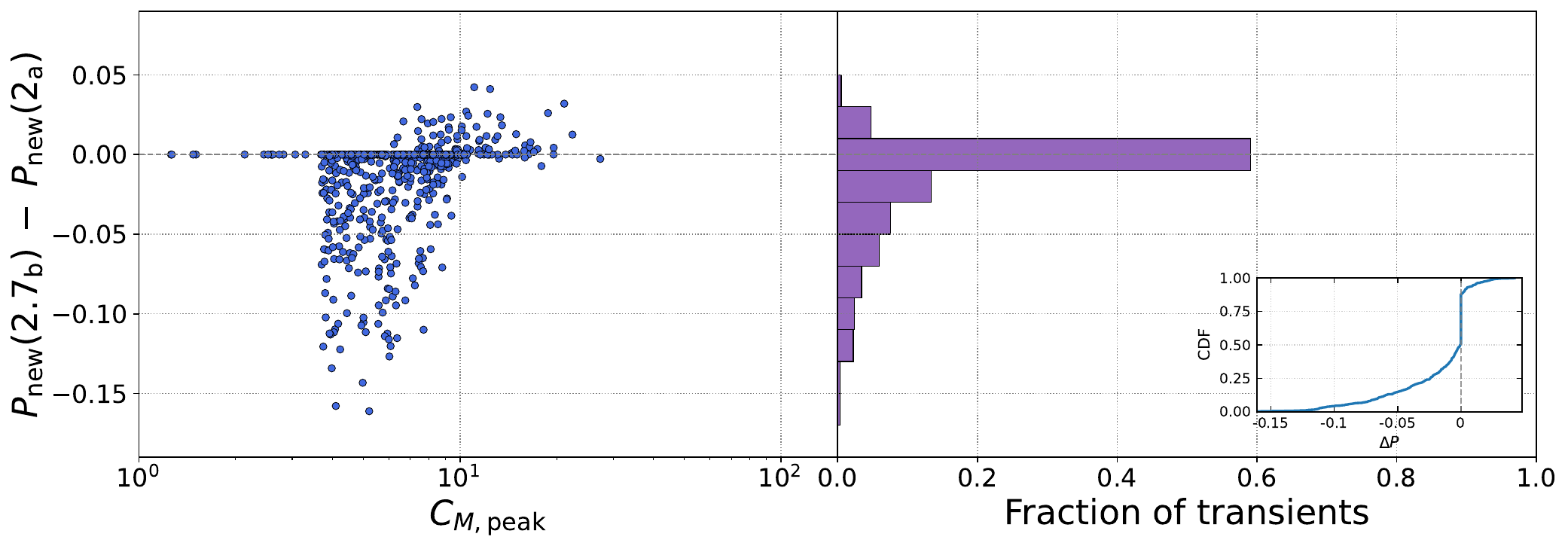}
   \caption{Effect of simultaneously accounting for seed-image exposure and peak--detection exposure differences.
\emph{Left:} $\Delta P=\pnew(2.7~\mathrm{ks},\ \mathrm{Case\ b})-\pnew(2~\mathrm{ks},\ \mathrm{Case\ a})$ as a function of measured peak counts, $C_{M,\mathrm{peak}}$.
\emph{Right:} histogram of $\Delta P$.
\emph{Inset:} cumulative distribution of $\Delta P$, showing the fraction of candidates with $|\Delta P|$ below a given value.}

    \label{fig:finalcompare}
\end{figure}

To explore the overall impact of the two factors just discussed, we compared the results for the 2.7~ks simulations analysed under Case~(b) and the nominal 2~ks simulations analysed under Case~(a) -- i.e. the two most discrepant cases. Figure~\ref{fig:finalcompare} shows that they are in very close agreement: for $\sim$60\% of candidates the difference in \pnew\ is consistent with zero, and for the remainder $|\Delta P| \lesssim 0.15$. This demonstrates that our inferred transient probabilities are robust to the main practical departures between our simulations and the LSXPS pipeline. 

\begin{figure}
    \centering
    \includegraphics[width=\linewidth]{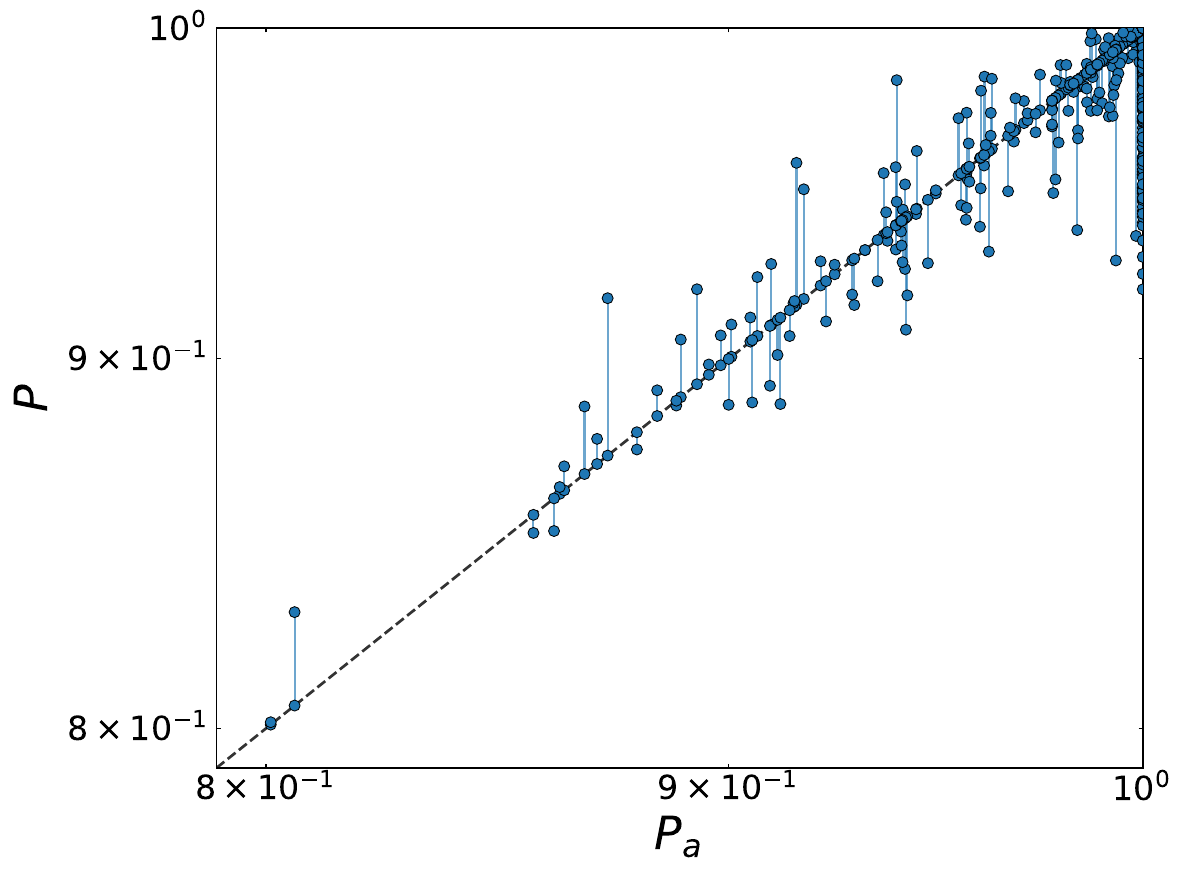}
    \caption{Candidate-by-candidate comparison of \pnew\ under the detection--peak exposure assumptions.
For each transient, the marker shows the Case~(a) value and the vertical bar spans the range between the Case~(a) and Case~(b) results, providing a practical estimate of the systematic uncertainty in \pnew\ associated with the choice of assumption.}
\label{fig:PP}

\end{figure}

Figure~\ref{fig:PP} summarises the detection-versus-peak exposure issue on a candidate-by-candidate basis. For each transient it compares the \pnew\ values obtained under Cases~(a) and~(b), with the vertical extent showing the spread between the two assumptions. This provides a practical estimate of the systematic uncertainty on \pnew\ for each candidate, and hence a natural way to report a probability range for new transient candidates in the live LSXPS system.

\section{Conclusions}

Using a Bayesian, simulation-based analysis we have been able to accurately reconstruct the probability distribution of the true intensity of sources detected by the \swift-XRT. This in turn allows us to robustly determine the probability that a given event is a transient, with effects such as the Eddington Bias corrected for. Using our approach we find that 632 objects serendipitously observed by \swift\ and identified by LSXPS as possible transients, but classified as likely statistical fluctuations, are in fact at least 2-$\sigma$ above the historical 3-$\sigma$ upper limit (500 of them at the more conservative 3-$\sigma$ confidence level used above). This technique thus enables a
much greater exploitation of the real- time transient detection capabilities of \swift\ and LSXPS.

These tools will shortly be added to the LSXPS pipeline as well as historical data, and the LSXPS website\footnote{\url{https://www.swift.ac.uk/LSXPS/transients}} will be revised to show the transient probabilities derived using this technique.

\section*{Acknowledgements}
This work made use of data supplied by the UK \emph{Swift} Science Data Centre at the University of Leicester. We acknowledge support from the UK Space Agency. The first author (S.\ Srivastava) thanks K.\ L.\ Page for early guidance during the first stages of this PhD, including practical training in DS9 and \swift/XRT data analysis.


\section*{DATA AVAILABILITY}

All of the original \emph{Swift} data used in this work are publicly available through the three \emph{Swift} data centres:
\url{https://www.swift.ac.uk/swift_live/},
\url{https://swift.gsfc.nasa.gov/archive}, and
\url{https://www.ssdc.asi.it/mmia/index.php?mission=swiftmastr}.
The transient probabilities and derived data products presented here are
available through the LSXPS catalogue web pages
(\url{https://www.swift.ac.uk/LSXPS/}) and can additionally be accessed
programmatically via the \texttt{swifttools} Python module, available
through \texttt{pip}.



\bibliographystyle{mnras}
\bibliography{example} 





\bsp	
\label{lastpage}
\end{document}